\documentclass[a4paper,11pt]{article}
\pdfoutput=2
\usepackage{jheppub} 
\usepackage[T1]{fontenc} 
\usepackage{subcaption}
\usepackage[utf8]{inputenc}
\newcommand{\RE}{\text{Re}}
\newcommand{\IM}{\text{Im}}

\title{Probing anomalous $Wtb$ couplings at the LHC in single top 
quark production through $t$-channel}

\author{Adil Jueid}

\affiliation[a]{INPAC, Shanghai Key Laboratory for Particle Physics and Cosmology,
Department of Physics and Astronomy, Shanghai Jiao Tong University, Shanghai 200240, China}
\affiliation[b]{D\'{e}partement de Math\'{e}matiques, 
Facult\'{e} des Sciences et Techniques,
Universit\'{e} Abdelmalek Essaadi, B. 416, Tangier, Morocco}
\emailAdd{adil.jueid@sjtu.edu.cn}

\abstract{ We study the sensitivity of certain observables to the anomalous right 
tensorial coupling in single top production at the LHC at $\sqrt{s}=13 \text{ TeV}$. 
The observables consist of asymmetries constructed from the energy and angles of the
decay products of the top  quark produced in single top production through 
$t$-channel. The computation is done at Leading Order (LO) and
Next-to-Leading Order (NLO) in the strong coupling in the $5$ flavor scheme. We have
estimated projected limits on the anomalous coupling, both at the parton level without
cuts and at the particle level with cuts. We find that the asymmetries are robust with 
respect to the higher order QCD corrections and are indeed a  very good probe of this 
anomalous coupling of the top. Hence they can be  be used as experimental probes of the same.}

\vspace{1cm}
\begin{document} 
\maketitle
\flushbottom 

%%%%%%%%%%%%%%%%%%%%%%%%%%%%
\section{Introduction}
\label{sec:introduction}
\begin{figure}[!t]
 \centering
 \includegraphics[width=0.48\linewidth]{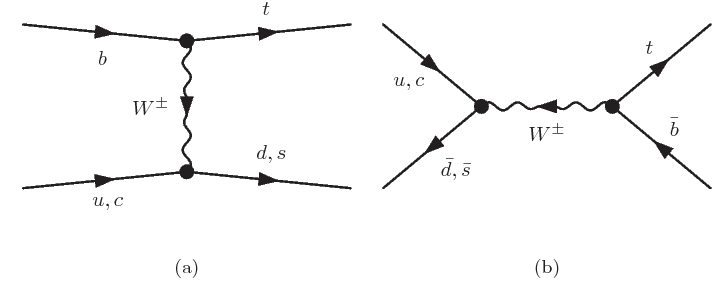}
 \hfill
  \includegraphics[width=0.48\linewidth]{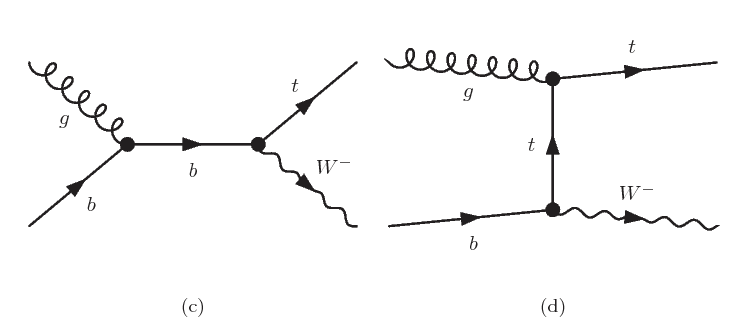}
 \caption{LO parton level Feynman diagrams of single top production at hadron colliders.}
 \label{diagrams-single}
\end{figure}

%%%%%%%%%%%%%%%%%%%%%%%%%%%%
Top quark is the heaviest among 
all the SM particles. This particle was discovered at the 
Tevatron-Fermilab by \texttt{CDF} \cite{Abe:1995hr} and
\texttt{D0} \cite{Abachi:1995iq} collaborations. It has 
a pole mass $m_t=173.1\pm0.6 \text{ GeV}$ 
\cite{Patrignani:2016xqp} which is very close to the electroweak 
symmetry breaking scale.
Due to its large mass, it can only be 
created at high energy experiments such as the Tevatron or the LHC with a reasonable number. 
 The top quark plays an important role in high energy physics as it 
is believed that, due its large mass, effects of new physics beyond 
the SM can be easily shown \cite{Beneke:2000hk, Han:2008xb, Bernreuther:2008ju}. 
Top quark is dominantly produced at the LHC, through QCD, in the pair
mode with a cross section approaching one nanobarn at 
$\sqrt{s} = 13 \text{ TeV}$. Due to the vector nature of the strong interaction, the produced 
top quark pairs are unpolarised. 
In addition to the pair production mode, 
top quark can be produced in association with a lighter
particle. This production mechanism proceeds through electroweak interaction. Hence it has smaller cross section, 
the maximum being $\sim 140$ pb.  However, the $V$-$A$ nature of the charged current interaction 
implies that the top quark produced in association are polarised. The much smaller cross section 
of the single top production along with the very large background from the top pair production 
meant that the first observation of single top production at Tevatron was made 14 years after 
the discovery of the top quark  \cite{Abazov:2009ii, Aaltonen:2009jj}. \\
There are three separate modes for single top production. They differ 
according to the associated particle produced with the top and the initial 
particles producing the top. These processes 
can be categorised according to the the virtuality of the $W$ boson. 
All these processes involve the $Wtb$ coupling; single top production through
$t$-channel (which has the largest cross section at the LHC), through $s$-channel 
and in association with a $W$-boson. 
The corresponding Feynman diagrams are depicted in Fig. \ref{diagrams-single}.

Single top production, although has smaller rate than $t\bar{t}$ production, is
phenomenologically very interesting. First, it allows a direct measurement of 
$V_{tb}$ in the Cabbibo-Kobayashi-Maskawa mixing matrix \cite{Alwall:2006bx, Cao:2015qta}. Inference on 
the $b$-quark density within the proton is possible as well by measuring
single top production cross section both through $t$-channel and $Wt$ process. 
In single top quark production, the 
produced top is highly polarised
allowing for a direct test of the $V-A$ structure of weak interaction
\cite{Mahlon:1999gz}. Finally, single top production is one of the
interesting channels to look for new physics beyond the Standard Model 
\cite{Tait:2000sh, Cao:2007ea, Cao:2006wk, Huang:2017yso}. Extensive studies of single
top production at hadron colliders including radiative corrections have been 
performed at NLO \cite{Campbell:2009ss, Bordes:1994ki, Stelzer:1997ns, 
Sullivan:2004ie, Campbell:2004ch, Campbell:2005bb, Cao:2004ap, Cao:2005pq, 
Bothmann:2017jfv} and NNLO \cite{Brucherseifer:2014ama, Berger:2016oht, 
Berger:2017zof} in the strong coupling. Furthermore, NLO calculations 
of single top production matched with parton showers become available 
within \textsf{MC@NLO} \cite{Frixione:2005vw, Frixione:2008yi} 
and \textsf{POWHEG} \cite{Alioli:2009je, Re:2010bp}. Recently, a transverse momentum
resummation at NLO+NLL for single top 
production through $t$-channel has been proposed in \cite{Cao:2018ntd}.
Single top production
cross sections have been measured by the ATLAS and CMS collaborations at
$\sqrt{s}=7\oplus8 \text{ TeV}$ \cite{Aad:2014fwa, Aad:2015upn, Aad:2015eto, 
Khachatryan:2014iya, Chatrchyan:2012ep, Chatrchyan:2011vp, Chatrchyan:2014tua, 
Chatrchyan:2012zca, Khachatryan:2016ewo} and $\sqrt{s}=13 \text{ TeV}$ 
\cite{Aaboud:2016lpj, Aaboud:2016ymp, Sirunyan:2016cdg}. These 
results were found in agreement with the SM predictions. \\

The top quark has a very short lifetime, 
$\tau_t \simeq G_F^{-1} m_t^{-3} \ll  m_t/\Lambda_\mathrm{QCD}^2$, which implies 
that it decays before hadronization effects take place. Hence, all its properties 
can be studied by looking at the kinematical distributions of 
its decay products. The 
top quark decays with almost a $100\%$ branching fraction into 
$W^\pm b$. Furthermore, due to the weak interaction universality, 
this process has certain pattern, the so-called $V-A$ structure which 
manifests itself at the lowest order in perturbation theory. Departures 
from this universal structure are possible through radiative 
corrections in the SM and/or new physics effects.
This departure might be seen as the presence of the so-called anomalous 
$Wtb$ couplings. One possible parametrization of these couplings is the
use of effective field theoretical approach where $SU(3)_c \otimes SU(2)_L \otimes U(1)_Y$ 
gauge invariant and 
$6$-dimensional operators are added to the SM Lagrangian 
\cite{Buchmuller:1985jz, Grzadkowski:2003tf, Grzadkowski:2010es, AguilarSaavedra:2008zc}. A global fit of these 
operators to the existing data has been recently 
done in \cite{Durieux:2014xla, Rosello:2015sck, Buckley:2015lku, Buckley:2015nca}. \\

The transition amplitude for top quark decay into a $W$-boson and 
a $b$-quark can be written as 

\begin{eqnarray}
  \mathcal{M}(t\to bW^+) = \frac{-e}{\sqrt{2} \sin \theta_W} \bar{u}_b(p_b) \Gamma^\mu_{tbW} u_t(p_t) \epsilon_\mu^*(q),
\label{anomalous}
\end{eqnarray}
with
$$
 \Gamma^\mu_{tbW} = - \frac{i g}{\sqrt{2}} \bigg[\gamma^\mu (V_L P_L + V_R P_R) + \frac{i}{M_W}
 \sigma^{\mu\nu} q_\nu (g_L P_L + g_R P_R) \bigg],
$$
 where $q = p_t - p_b$ is the momentum of the $W^\pm$ boson (which is assumed to be on-shell)
 and $P_{L,R}=1/2(1\pm\gamma_5)$ are the projection operators, 
 $V_L, V_R, g_L \text{ and } g_R$ are called anomalous couplings.
 In the SM, at tree level, $V_L=V_{tb}$ and $V_R=g_L=g_R=0$ whereas 
 EW and QCD radiative corrections induce non-zero values of the anomalous couplings. 
 Computations of the anomalous couplings have been performed in 
 \cite{Czarnecki:1990kv, Li:1990qf, Bernreuther:2008us, Drobnak:2010ej, GonzalezSprinberg:2011kx, 
 Duarte:2013zfa, Gonzalez-Sprinberg:2015dea, Arhrib:2016vts, Ayala:2016djv, Martinez:2016fyd} 
 both in the SM and certain extensions of it. 
 It was found that SM corrections to anomalous couplings are extremely small, i.e 
 $g_L=-(1.247 +0.002747 i)\times 10^{-3}$, $g_R=-(8.6 + 2.05 i) \times 10^{-3}$
 and $V_R=(2.911 + 0.9 i) \times 10^{-3}$ \cite{Arhrib:2016vts}.
 Furthermore, the dominant contribution comes from QCD  
 whereas EW corrections are subleading accounting about $8$--$15\%$ of the total contribution.
 Contributions of the anomalous $Wtb$ couplings to various flavour 
 observables have been considered in \cite{Drobnak:2011wj, Drobnak:2011aa}. \\

As mentioned above, the top quark produced in association with a quark or $W$, is highly polarised. 
This polarisation is decided by  the anomalous $tbW$ couplings involved in the production of the single top. 
The production cross-section, the polarisation of the top and the energy distributions of the decay products 
of the single top, all carry information about these anomalous couplings.
In fact, recently a study \cite{Godbole:2006tq} had shown how one can 
simultaneously study the top polarisation as well as 
the $Wtb$ anomalous coupling and constructed some asymmetries in the observables 
which are sensitive to both. However, it did not make any reference to a specific top production mechanism. 
It was pointed out in \cite{Cao:2015doa} that measuring accurately single top production cross sections 
throught $t$- and $tW$ channels would constrain effectively 
the anomalous $Wtb$ couplings. In this paper we wish extend the analysis 
of the observables suggested in \cite{Godbole:2006tq} as  well as 
the cross-section information, to the single top production via $t$-channel.

Thus the aim of this paper is to make a study of the sensitivity of the LHC  to the anomalous coupling 
$g_R$ using energy and angular observables in $t$-channel single top production at 
$\sqrt{s}=13$ TeV. The observables used in our analysis consist of asymmetries based on energy and angular 
distributions of the top quark decay products. 
As we will see in this paper, their use will give additional information about the existence of possible 
anomalous couplings in the $Wtb$ vertex. Since these observables depend on polarisation, it is expected
that they will be robust against QCD radiative corrections. In fact this was demonstrated explicitly in 
an analysis of the charged Higgs production \cite{Godbole:2011vw}. 
In this work also we perform a comparative study at LO and NLO in the strong coupling constant. 

We use observables suggested in \cite{Prasath:2014mfa}, but many of them  were proposed long time ago 
in \cite{Godbole:2006tq,Shelton:2008nq} and were used for several studies 
(see \cite{Rindani:2011pk,Godbole:2011vw,Belanger:2012tm, Godbole:2015bda, Rindani:2015vya}). 
In addition, we separate the study into
two different categories; parton and particle level. In the former case, no cuts are imposed 
on the particles' momenta while in the latter loose cuts are imposed. The effects of 
such kinematical cuts were found to be quite important because these affect 
the shape of the distributions and hence also the projected limits from the results at the parton level.
 We stress out that these 
asymmetries are of extreme importance for future experimental
analyses and can be used in several channels that involve the top quark and in different 
collider machines. We will show, in this paper, that using NLO matrix elements is mandatory 
for future experimental analyses. In this study, following the limits from $\text{BR}(b \to s \gamma)$, we set
$V_R=g_L=0$ and $g_R$ will be taken to be real.\\

The paper is outlined as follows: In section \ref{sec:limits}, we review the limits on the 
anomalous $Wtb$ couplings from direct experimental searches and from statistical fits to several 
measurements. The setup of the calculation and  details about event selection are summarized in 
section \ref{sec:setup}. In section \ref{sec:single-top}, we show the computations of single top 
production through $t$-channel at LO and NLO both in the SM and the SM augmented by anomalous 
$Wtb$ coupling. The studied observables are briefly discussed 
in section \ref{sec:observables}. Numerical results 
and sensitivity projections are shown in section \ref{sec:results}. In section \ref{sec:conclusions}, we
draw our conclusions. Details about the interpolation procedure are 
shown in appendix \ref{sec:appen1}.

%%%%%%%%%%%%%%%%%%%%%%%%%%%%%%%%%%%%%%%%%%%%%%%%%%%%%%%
\section{Limits on anomalous $Wtb$ couplings}
\label{sec:limits}
%%%%%%%%%%%%%%%%%%%%%%%%%%%%%%%%%%%%%%%%%%%%%%%%%%%%%%%

Anomalous $Wtb$ couplings are constrained both {\it indirectly} from flavour changing decays 
such as $b \rightarrow s + \gamma$ 
as well as from measurements at both the Tevatron and the LHC: those of helicity 
fractions of the $W$ produced in top decay, both in pair production of the top as 
well as the single top production, as well as the spin-spin correlations in top pair production etc.
A summary of  early constraints can be found in \cite{Bernreuther:2008ju}.  
Measurement of $b \rightarrow s + \gamma$ branching ratio 
constrains $V_{R}, g_{L}$ strongly, however $g_{R}$ is rather weekly constrained \cite{Grzadkowski:2008mf}. 
Measurements of the $W$-boson helicity fraction in top pair production 
at the Tevatron \cite{Abazov:2012uga} as well at the LHC \cite{Aad:2012ky, Chatrchyan:2013jna, Aaboud:2016hsq}. 
In fact an analysis of a combination of the measurements of 
the BR($b \rightarrow s \gamma$) \cite{Grzadkowski:2008mf} and the $W$ helicity 
fractions at the Tevatron \cite{Abazov:2012uga} together, had  shown that 
$|g_{R}|$ is the only coupling that could have nontrivial values. The large increase of 
single top production cross-sections with increasing energy, from Tevatron to the LHC,
meant that one could also use the single top processes to 
this end as well \cite{Chatrchyan:2013jna, Khachatryan:2014vma, Aad:2015yem, Khachatryan:2016sib, Aaboud:2017aqp}.  
LHC experiments have used helicity fraction of the $W$ produced in the $t$ decay \cite{Aad:2012ky}, 
the double differential decay rate of the singly produced top quark \cite{Aad:2015yem}, 
asymmetries constructed out of $W$-boson angular distributions \cite{Aaboud:2017aqp}
as well as the triple differential decay distributions 
for the $t$ quark \cite{Aaboud:2017yqf}. In these analyses limits on various anomalous couplings are obtained 
under different assumptions; sometimes letting all the couplings vary around their SM values 
or sometimes keeping some of the couplings fixed at their SM values and so on. Furthermore, depending on 
the variables used, limits can be obtained on the real or imaginary parts of these anomalous 
couplings.  Analyses which tried to constrain all the anomalous couplings simultaneously, using only 
the data on single top at the LHC \cite{Khachatryan:2016sib,Aaboud:2017yqf} with no 
assumptions on  value of $V_{L}$ still allow values of $|g_{R}| \sim 0.1$--$0.2$. 
Fits assuming $V_{L} =1$ from $t \bar t$ production \cite{Aaboud:2016hsq} by ATLAS also allow large 
values of $g_{R} (\sim 0.7)$ but these are in conflict with the cross-section measurements of single top 
processes and hence cannot be taken seriously. 
On the other hand, a phenomenological analysis of only the collider data, viz. the single top 
cross-sections and $W$ helicity fractions from both the Tevatron 
and the LHC \cite{Fabbrichesi:2014wva}, obtains mildest 
constraints on $|g_{R}|$ and $V_{R}$. \\

In addition, a combination of different measurements corresponding to  electron 
and neutron electric dipole  moments (EDM), top quark observables and oblique 
parameters, constrains both the imaginary and real parts of the tensorial couplings $g_{R}$ 
and $g_{L}$ \cite{Cirigliano:2016njn, Cirigliano:2016nyn}.
In the end we note that while the imaginary parts of anomalous tensorial 
couplings are most severely constrained by EDM's of neutron and electron,
BR($b \rightarrow s \gamma$) constrains $|g_{L}|$ and $W$-helicity fraction 
from the collider data constrain the $|g_{R}|$ most effectively.  Recentl global fits 
to the LHC and Tevatron data \cite{Birman:2016jhg, Castro:2016jjv, Hioki:2016xtc, Deliot:2017byp, Xiong:2018mkc}  
and the limits on $g_{R}$ are not very strong.

A summary of current limits and analyses can be found in \cite{Patrignani:2016xqp}. For reference, 
we depict some limits from 
experimental searches and from BR($b\to s\gamma$) in Table \ref{limits-LHC}. All this discussion 
thus tells us that it is of great interest to see how the collider data on cross-section and 
top polarisation in single top production can constrain $g_{R}$. We proceed now to discuss the procedure to do so using 
new observables.   

\begin{table}[!t]
 \begin{center}
  \begin{tabular}{c | c | c}
   \hline \hline
   Constraint & Limits & Reference \\ \hline \hline
   BR($b\to s\gamma$)  & $-0.15 < \RE(g_R) < 0.57, -7 \times 10^{-4} \leq V_R \leq 2.5 \times 10^{-3}$ & \cite{Grzadkowski:2008mf} \\ 
                       & $-1.3 \times 10^{-3} \leq g_L \leq 4 \times 10^{-4}$                          &     \\ \hline
   $W$ helicity fractions &  $\textrm{Re}(V_R) \in [-0.20,0.23], \textrm{Re}(g_L) \in [-0.14,0.11],$ & \cite{Aad:2012ky} \\
                          &  $\textrm{Re}(g_R) \in [-0.08,0.04].$                                    & \\ 
                          &  $\RE(g_R) \in [-0.24,0.20],  \RE(g_L) \in [-0.14,0.10]$                 & \cite{Chatrchyan:2013jna} \\ 
                          &  $\RE(g_L) \in [-0.30, 0.25], \RE(g_R) \in [-0.15, 0.10]$                & \cite{Khachatryan:2014vma} \\ \hline
double differential cross section & $\RE\left(\frac{g_R}{V_L}\right) \in [-0.36,0.10] 
\IM\left(\frac{g_R}{V_L}\right) \in [-0.17,0.23]$ & \cite{Aad:2015yem} \\ \hline                           
$t$-channel cross section &  $V_L > 0.98, |V_R| < 0.16$ & \cite{Khachatryan:2016sib} \\
                          &  $|g_L| < 0.057, -0.049 < g_R < 0.048$ & \\ \hline
$W$-boson polarisation    &  $ V_R \in [-0.24, 0.31], g_L \in [-0.14, 0.11],$ & \cite{Aaboud:2016hsq} \\
                          &  $ g_R \in [-0.02, 0.06]\cup[0.74,0.78]$ & \\ 
                          &  $\IM(g_R) \in [-0.18,0.06]$  & \cite{Aaboud:2017aqp} \\ \hline
triple differential cross section &   $\left|\frac{V_R}{V_L}\right| < 0.37, \left|\frac{g_L}{V_L}\right| < 0.29$ & \cite{Aaboud:2017yqf} \\
                           & $ \RE\left(\frac{g_R}{V_L}\right) \in [-0.12,0.17], \IM\left(\frac{g_R}{V_L}\right) \in [-0.07,0.07]$ & \\ \hline \hline
\end{tabular}
 \end{center}
 \caption{Summary of the limits on anomalous $Wtb$ couplings from
 BR($b\to s\gamma$) and from direct experimental searches. A complete listing can 
 be found in the \textbf{PDG} \cite{Patrignani:2016xqp}.}
 \label{limits-LHC}
\end{table}

%%%%%%%%%%%%%%%%%%%%%%%%%%%%%%%%%%%%%%%%%%%%%%%%%%%%%%%%
%%%%%%%%%%%%%%%%%%%%%%%%%%%%%%%%%%%%%%%%%%%%%%%%%%%%%%%%
\section{Setup and event selection}
\label{sec:setup}
%%%%%%%%%%%%%%%%%%%%%%%%%%%%%%%%%%%%%%%%%%%%%%%%%%%%%%%%
%%%%%%%%%%%%%%%%%%%%%%%%%%%%%%%%%%%%%%%%%%%%%%%%%%%%%%%%

For this study, we consider single top production
through $t$-channel in $p p$ collisions at $\sqrt{s} = 13 \text{ TeV}$.
For electroweak couplings, we use the 
$G_\mu$-scheme, in which the input parameters are $G_F, \alpha_\textrm{em}(0) \text{ and } M_Z$. 
We choose $G_F = 1.16639 \times 10^{-5} \text{GeV}^{-2}$, $\alpha_\textrm{em}^{-1}(0)=137$ and
$M_Z=91.188 \text{ GeV}$. From these input parameters, $M_W$ and $\sin^2\theta_W$ are obtained.
Furthermore, the top quark pole
mass is chosen to be $m_t=173.21 \text{ GeV}$. The computation of single top production
cross section was done in the $5$FS with massive (massless) $b$-quark at LO (NLO).
We use the
\textsf{NNPDF30} PDF sets \cite{Ball:2014uwa} 
with the \textsf{LHAPDF6} interpolator tool \cite{Buckley:2014ana}
with $\alpha_s(M_Z)=0.118$. Throughout this study, we will use fixed factorization 
and renormalization scales, i.e $\mu_F=\mu_R=m_t$. 

Events are generated with \textsf{Madgraph5$\_$aMC@NLO} \cite{Alwall:2011uj, Alwall:2014hca} 
in the SM at Leading Order (LO) and Next-to-Leading Order (NLO) 
and the SM with anomalous couplings at LO. The right tensorial
anomalous coupling $g_R$ is implemented by hand in a \textsf{UFO} model file
\cite{Degrande:2011ua}. The model file were validated by comparing some calculations to 
several results existing in the literature \cite{Rindani:2011pk, Cao:2015doa} concerning 
cross section calculations and several distributions in $t$- and $tW$ channels and
we found excellent agreement.
The produced events were decayed with \textsf{MadSpin} \cite{Artoisenet:2012st} which uses the method developed
in \cite{Frixione:2007zp} to keep full spin correlations. 
The decayed events are passed to \textsf{PYTHIA8} \cite{Sjostrand:2014zea} 
to include parton showers (ISR and FSR) and 
hadronization. Parton showering algorithm is based on dipole type $p_\perp$ evolution 
\cite{Sjostrand:2004ef}. The other parameters are set according to the \texttt{Monash} 
tune \cite{Skands:2014pea}. We have adopted the \textsf{MC@NLO} 
scheme \cite{Frixione:2002ik} for consistent matching of hard-scattering matrix elements
and parton shower MC. \textsf{RIVET} \cite{Buckley:2010ar} was used 
for analysis of the events. For jet clustering,
we use \textsf{FastJet} \cite{Cacciari:2011ma} with an anti-$k_\perp$ algorithm and jet radius
$R=0.4$ \cite{Cacciari:2008gp}. \\

We, first, perform a partonic level analysis (no showers,
no soft QCD effects and no cuts on the kinematical quantities). We then, perform a 
particle level analysis (without detector effects) of the showered events. 
Throughout this paper, we will show
results at both the partonic and the particle levels. For the particle level analysis,
we require a topology consisting of exactly one isolated
charged lepton (electron or muon), missing energy 
$E_{\text{T}}^{\text{miss}}$ 
and at least two jets with at least one of them is $b$-tagged. First, 
we require exactly one isolated charged lepton with transverse momentum 
$p_\perp(\text{lepton}) > 10 \text{ GeV}$ and pseudorapidity $|\eta| < 2.5$. 
We require at least two jets where one of them is tagged with $|\eta| < 2.5$ and 
$p_\perp(\text{jet}) \geq 25 \text{ GeV}$. Further isolation requirements
are applied to jets, i.e the angular separation should be always 
$\Delta R = \sqrt{\Delta\eta^2 + \Delta\phi^2} > 0.5$
for any two jets in the event and $\Delta R(\text{jet}, \text{lepton})>0.4$. 

%%%%%%%%%%%%%%%%%%%%%%%%%%%%%%%%%%%%%%%%%%%%%%%%%%%%%
\section{Single top production through $t$-channel}
\label{sec:single-top}
%%%%%%%%%%%%%%%%%%%%%%%%%%%%%%%%%%%%%%%%%%%%%%%%%%%%%

In this section, we discuss single top production cross
section in the SM at LO and NLO. We illustrate at the end of the 
section the method that we used to include anomalous $Wtb$ coupling in the production
at NLO.

%%%%%%%%%%%%%%%%%%%%%%%%%%%%%%%%%%%%%%%%%%%%
\subsection{LO calculation in the SM}
%%%%%%%%%%%%%%%%%%%%%%%%%%%%%%%%%%%%%%%%%%%%
At LO, there are two generic contributions to the $t$-channel process. 
The first contribution corresponds to the subprocess:
\begin{eqnarray}
b \hspace{0.2cm} q \to t \hspace{0.2cm} q',
\label{diag-1}
\end{eqnarray}
and the second contribution represents the subprocess:
\begin{eqnarray}
 b  \hspace{0.2cm} \bar{q}' \to t \hspace{0.2cm} \bar{q},
 \label{diag-2}
\end{eqnarray}
where $q=u,c$ and $q'=d,s$. Furthermore, contributions 
to the $t$-channel process involving the negligble elements of the Cabbibo-Kobayashi-Maskawa (CKM)
mixing matrix such as $V_{td} \text{ and } V_{ts}$ were not taken into account.
We have computed the inclusive LO cross sections at 
$\sqrt{s} = 13$ TeV. Due to the dominance 
of the valence u-quark PDF over the sea anti-quarks, 
the subprocess \ref{diag-1} gives the dominant contribution
which accounts of about $77 \%$ of the total cross
section.
%%%%%%%%%%%%%%%%%%%%%%%%%%%%%%%%%%%%%%%%%%%%%%%%
\subsection{The $t$-channel at NLO in the SM}
%%%%%%%%%%%%%%%%%%%%%%%%%%%%%%%%%%%%%%%%%%%%%%%%
\begin{figure}[!t]
\begin{center}
\includegraphics[width=12cm, height=4.5cm]{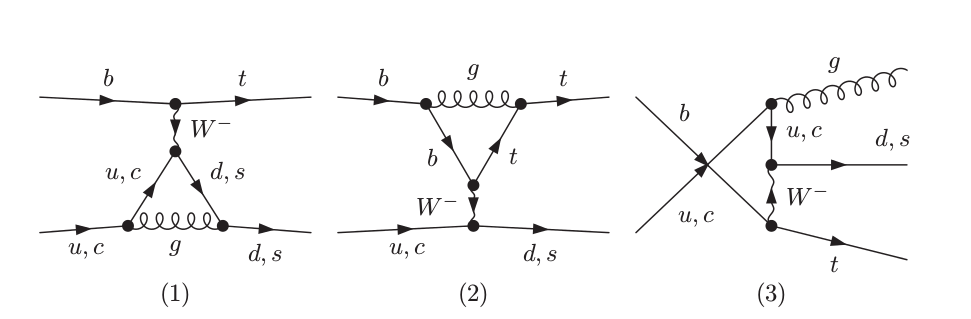}
\vspace{-2cm}
\includegraphics[width=12cm, height=4.5cm]{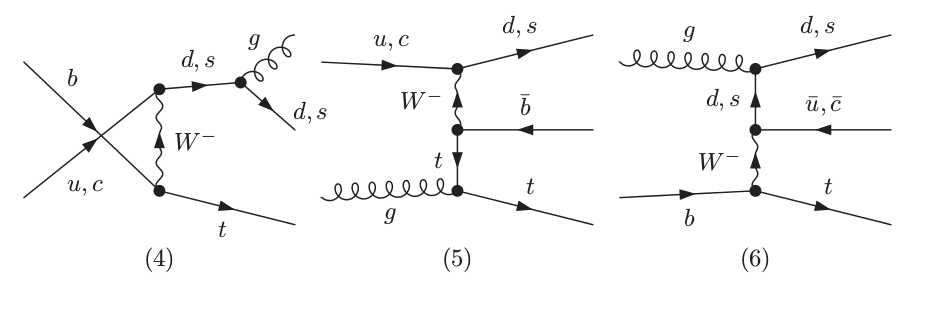}
\vspace{0.7cm}
\end{center}
\caption{Feynman diagrams contributing to $b q \to t q'$ subprocess at NLO in the SM where $q=u,c$ and
$q'=d,s$.}
\label{diagrams-1}
\end{figure}
\begin{figure}[!t]
\begin{center}
\includegraphics[width=12cm, height=4.5cm]{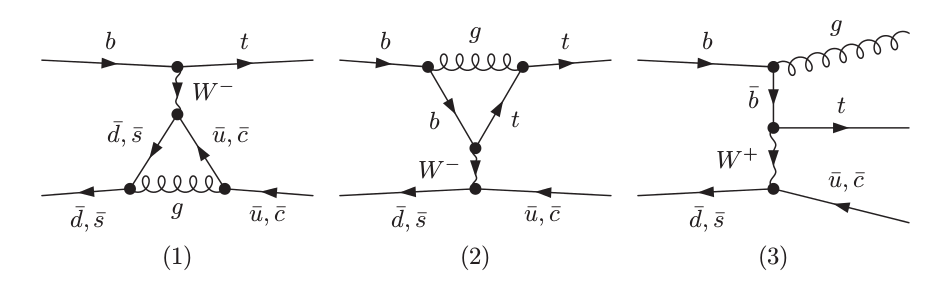}
\vspace{-2cm}
\includegraphics[width=12cm, height=4.5cm]{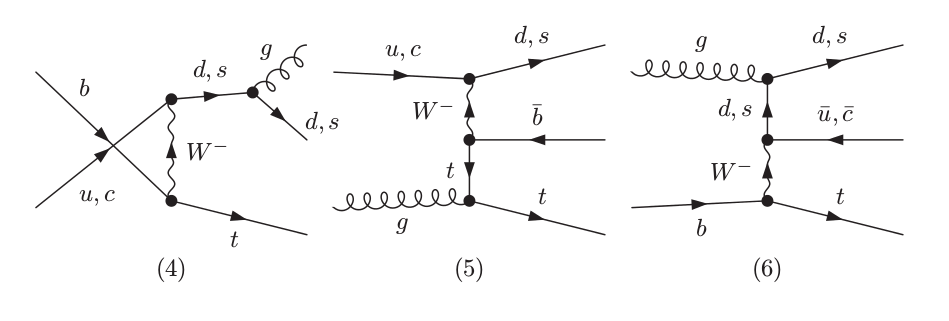}
\vspace{0.7cm}
\end{center}
\caption{Feynman diagrams contributing to $b \bar{q}' \to t \bar{q}$ subprocess at NLO in the SM where $q=u,c$ and
$q'=d,s$.}
\label{diagrams-2}
\end{figure}
Further contributions to the $t$-channel process, at NLO exist. Such contributions
include virtual
one-loop corrections to the tree level process as well as tree level $2 \to 3$ real emission processes
where the additional emitted parton is soft or collinear. 
Parton level Feynman diagrams are depicted in figures \ref{diagrams-1} and \ref{diagrams-2}. All the 
possible flavours that might contribute to this process were included. Due to color
conservation, box diagrams do not contribute to the cross section at NLO. 

 \begin{table}[!t]
  \begin{center}
  \begin{tabular}{l | l | l | l }
   \hline \hline
   Order  &            \hspace{0.5cm} $\sigma$ [pb] \hspace{0.5cm} & \hspace{0.5cm} $\delta\sigma_\mu$ [\%] &  \hspace{0.5cm} $\delta\sigma_\textrm{PDF}$ [\%] \\ \hline
   LO  \hspace{1cm}  & \hspace{0.5cm} $128.67$ \hspace{0.5cm}& \hspace{0.5cm} $^{+9.12}_{-11.3}$ & \hspace{0.5cm} $\pm8.88$    \\ 
                     & \hspace{0.5cm} $50.50$ \hspace{0.5cm}& \hspace{0.5cm} $^{+10.4}_{-12.5}$ & \hspace{0.5cm} $\pm9.65$ \\
   \hline \hline
   NLO  \hspace{1cm} & \hspace{0.5cm} $141.8$ \hspace{0.5cm} & \hspace{0.5cm} $^{+2.8}_{-2.5}$ & \hspace{0.5cm}  $\pm1.2$  \\
                     & \hspace{0.5cm} $75.11$ \hspace{0.5cm} & \hspace{0.5cm} $^{+2.8}_{-2.9}$ & \hspace{0.5cm} $\pm1.1$ \hspace{0.5cm} \\
   \hline 
   \hline
  \end{tabular}
  \end{center}
  \caption{$t$-channel production cross 
  section at LO and NLO at $\sqrt{s}=13$ TeV. Uncertainties due to scale variations and 
  PDF are shown. First rows show the values of the cross section without cuts and in the 
  second rows the cross sections are computed using the cuts highlighted in section \ref{sec:setup}.}
  \label{xsec-LO-NLO}
 \end{table}

We have computed the total cross section at $\sqrt{s}=13 \text{ TeV}$ with 
$\mu_R=\mu_F=m_t$. The dominance of the
valence $u$-quark PDF implies that the 
cross section from the contribution of Feynman diagrams in Fig. \ref{diagrams-1} is 
dominant. We have also estimated theoretical uncertainties on the inclusive 
cross section both at LO and NLO from scale variations and from
PDF. Theoretical uncertainties which are due
to scale variations are estimated by varying 
simultaneously the hard factorization and renormalization scales around
their nominal values, i.e :
\begin{eqnarray}
 0.5 \leq \mu_{R}/\mu_{R,0}, \mu_{F}/\mu_{F,0}  \leq 2,
\end{eqnarray}
where $\mu_{R,0}$ ($\mu_{F,0}$) is the central renormalization (factorization) scale. Thus, obtaining
an envelope of the nine possible variations. PDF uncertainties are obtained using the replicas
method where each PDF set has one central and 50 members corresponding to the minimal fit
and the eigenvectors respectively. 
The results for the inclusive cross section at LO and NLO along 
with their theoretical uncertainties are depicted in Table \ref{xsec-LO-NLO}. 
After cuts were imposed, the total rate decreases by a factor of $\simeq 2.5$ 
and $\simeq 1.9$ at LO and NLO respectively. Furthermore, theoretical
uncertainties due to scale variations increase in both LO as well as NLO 
with the former has even larger relative increase than the latter. The most 
important consequence of imposing cuts on the decay products is that
the $K$-factor of the fiducial cross section increase to about $1.5$ while 
the theoretical uncertainties at LO
do not change much. An immediate consequence of such observation is that 
the fiducial cross section at NLO is outside the allowed range of scale 
variations' uncertainty at LO since 
$\sigma_\textrm{LO}^\textrm{max}=50.5+\delta\sigma_\mu^+\oplus\delta\sigma_\textrm{PDF}^+\approx 58$ pb $< \sigma_\textrm{NLO}$.
Hence, for any future analysis or fit involving the cross section of single top through $t$-channel, at least the calculation at NLO
should be used. 

%%%%%%%%%%%%%%%%%%%%%%%%%%%%%%%%%%%%%%%%%%%%%%%%%%%%%%%
\subsection{Including anomalous $Wtb$ couplings}
%%%%%%%%%%%%%%%%%%%%%%%%%%%%%%%%%%%%%%%%%%%%%%%%%%%%%%%

The presence of anomalous $Wtb$ couplings modifies single 
top quark production cross 
sections. The production cross section of a top quark 
through $t$-channel in $pp$ collisions can be expressed 
as function of anomalous right tensorial coupling $g_R$ as
\begin{eqnarray}
 \sigma_{\text{t-ch}} = \sigma_{\text{t-ch}}^{\text{SM}} \bigg(1 + \kappa_1 g_R + 
 \kappa_2 |g_R|^2   \bigg),
 \label{sig-Anom}
\end{eqnarray}
where $\kappa_{1,2} \equiv \kappa_{1,2}(m_t,m_b,M_W,\sqrt{s})$. 
They are determined 
from a fit and are given by
\begin{eqnarray*}
\kappa_1 = 0.45485 \quad (0.66433) \qquad \textrm{without cuts (with cuts)}, \\
\kappa_2 = 2.05348 \quad (3.21011) \qquad \textrm{without cuts (with cuts)},
\end{eqnarray*}
at $\sqrt{s}=13$ TeV. In eqn. \ref{sig-Anom}, $\sigma_{\text{t-ch}}^{\text{SM}}$ 
is the SM cross section at LO (see e.g. Table \ref{xsec-LO-NLO}).
We can see that imposing cuts strengthen the dependence of the cross
section on the anomalous coupling by 
a factor of $\simeq 1.5$. Although the quadratic 
term is about 5 times larger than the linear one 
its contribution to the cross section is mild even for the extreme values
of $g_R$, i.e $g_R=\pm0.2$. The results we obtained were compared with 
those presented in \cite{Cao:2015doa} and we found excellent agreement. \\

Taking into account the anomalous $Wtb$ coupling in the production at NLO is not 
straightforward. The reason is that one cannot renormalize high dimensional operators using 
the traditional on-shell renormalization schemes. Using alternative schemes for the renormalization
of the SM wave functions and parameters will result in large theoretical uncertainties.
However, including anomalous $Wtb$ couplings in the production is very interesting since they 
completely change the chiral structure of the $Wtb$ vertex and hence the top quark polarisation
which in fact improve the sensitivity of most the observables on the anomalous couplings. 
We include their effects as a shift on the production while neglecting interference between the virtual 
corrections and tree level amplitudes with anomalous couplings.
Hence, three samples will be generated with the first one corresponds to SM amplitude
at NLO, the second to the LO amplitude in the SM and the third one to the amplitude with anomalous couplings 
at tree level. The transition amplitude for the process  
$$
p p  \to t + X \to b \ell^+ \nu_\ell + X,
$$
can be written as
\begin{eqnarray}
 \mathcal{M}(\lambda) = \mathcal{P}(\lambda) \mathcal{D}(\lambda),
 \label{amplitude}
\end{eqnarray}
where $\mathcal{P}$ ($\mathcal{D}$) is the production (decay) matrix elements 
for the top quark. 

For convenience ($\lambda$) and ($\lambda'$) stand for the 
helicity labelling of all particles.
The pure SM tree level amplitude (equivalent to $g_R = 0$) is $\mathcal{P}_0^\text{SM}$. The
contribution of the anomalous coupling $g_R$ will only be taken into account at tree-
level and will be denoted $\mathcal{P}_0^{g_R}$. The full tree-level amplitude, anomalous
amplitude, will be denoted $\mathcal{P}_0^\text{ano.}$, such that
\begin{eqnarray}
 \mathcal{P}_0^\text{ano.} = \mathcal{P}_0^\text{SM} + \mathcal{P}_0^{g_R}
\end{eqnarray}
Since the radiative corrections at the level of the decay are very small in
the SM  (see e.g. \cite{Gao:2012ja}), the decay part will be considered at tree-level only.
For the production part, at tree-level, with the inclusion of the anomalous
part we have
\begin{eqnarray}
 \mathcal{P}_\textrm{tree}(\lambda,\lambda') &=&  \mathcal{P}_\textrm{LO}(\lambda,\lambda') = 
 (\mathcal{P}_0^\text{SM}(\lambda) + \mathcal{P}_0^{g_R}(\lambda')) 
 (\mathcal{P}_0^\text{SM}(\lambda) + \mathcal{P}_0^{g_R}(\lambda'))^* \nonumber \\
 &\equiv&  \mathcal{P}_0^\text{ano.}(\lambda) \mathcal{P}_0^\text{ano.}(\lambda')^*.
\end{eqnarray}
The one-loop radiative corrections will only apply to the SM part. The
first higher order contribution consists of the one-loop $2\to2$ virtual correction 
that includes also counterterms, $\delta \mathcal{P}_V(\lambda)$. To this one needs to add
the pure SM radiative, $2 \to 3$ emission. First of all, for the $2 \to 2$ one-loop
virtual correction and the $g_R$ contribution we may write
\begin{eqnarray}
 \mathcal{P}_1(\lambda,\lambda') &=&  
 (\mathcal{P}_0^\text{SM}(\lambda) + \delta \mathcal{P}_V(\lambda) + \mathcal{P}_0^{g_R}(\lambda')) 
 (\mathcal{P}_0^\text{SM}(\lambda) + \delta \mathcal{P}_V(\lambda) + \mathcal{P}_0^{g_R}(\lambda'))^* \nonumber \\
 &\simeq& \mathcal{P}_V(\lambda,\lambda') + \mathcal{P}_0^\text{ano.}(\lambda) \mathcal{P}_0^\text{ano.}(\lambda')^*
 - \mathcal{P}_0^\text{SM}(\lambda) \mathcal{P}_0^\text{SM}(\lambda')^*,
\end{eqnarray}
where
\begin{eqnarray}
 \mathcal{P}_V(\lambda,\lambda') = \mathcal{P}_0^\text{SM}(\lambda) \mathcal{P}_0^\text{SM}(\lambda')^* +
 \mathcal{P}_0^\text{SM}(\lambda) \delta \mathcal{P}_V(\lambda')^* + 
  \delta \mathcal{P}_V(\lambda) \mathcal{P}_0^\text{SM}(\lambda')^*
  \label{decom2}
\end{eqnarray}
Including and integrating over real emission, $\mathcal{P}_R^\text{SM}(\lambda)$, and adding it to
the virtual correction $\mathcal{P}_V(\lambda,\lambda')$ of Eq. \ref{decom2} 
will give the full NLO SM result. What we will refer to as the full NLO (including the 
LO anomalous part) is
\begin{eqnarray}
 \mathcal{P}_\text{NLO}(\lambda,\lambda') = \underbrace{\mathcal{P}_V(\lambda,\lambda') + 
 \mathcal{P}_R^\text{SM}(\lambda,\lambda')}_\text{sample 1}
 + \underbrace{\mathcal{P}_0^\text{ano.}(\lambda) \mathcal{P}_0^\text{ano.}(\lambda')^*}_\text{sample 2}
 - \underbrace{\mathcal{P}_0^\text{SM}(\lambda) \mathcal{P}_0^\text{SM}(\lambda')^*}_\text{sample 3}.
\end{eqnarray}
In order to reproduce the data including both the NLO SM and the
anomalous contribution (with its quadratic part) we therefore, for the same
phase space point, generate three samples. One for the full anomalous part
at tree-level, one for the SM tree-level part (which has to be subtracted to
avoid double counting) and one for the SM NLO (which includes the SM
tree, virtual and real emission).
%===============================
\section{Observables}
\label{sec:observables}
%===============================
In this section, we review the observables that we will be using for our analysis
of anomalous $Wtb$ couplings. They consist of asymmetries constructed from the 
energy and angular distributions of the top quark decay products
in single top production through $t$-channel. \\

We define an asymmetry with respect to a kinematical 
variable $\mathcal{O}$ by:
\begin{eqnarray}
 A_{\mathcal{O}} = \frac{\int_{\mathcal{O}_\text{min}}^{\mathcal{O}_\text{R}} \frac{d\sigma}{d\mathcal{O}} d\mathcal{O}
 - \int_{\mathcal{O}_\text{R}}^{\mathcal{O}_\text{max}} \frac{d\sigma}{d\mathcal{O}} d\mathcal{O}}
 {\int_{\mathcal{O}_\text{min}}^{\mathcal{O}_\text{R}} \frac{d\sigma}{d\mathcal{O}} d\mathcal{O} + 
 \int_{\mathcal{O}_\text{R}}^{\mathcal{O}_\text{max}} \frac{d\sigma}{d\mathcal{O}} d\mathcal{O}},
\end{eqnarray}
where $\frac{d\sigma}{d\mathcal{O}} = \frac{d\sigma(pp\to t X\to \ell^+\nu_\ell b X)}{d\mathcal{O}}$,
$\ell=e,\mu$, is the differential
cross section of the top quark with respect to the variable $\mathcal{O}$ and $\mathcal{O}_R$ is a reference point
around which the asymmetry will be evaluated. $\mathcal{O}_R$ will be chosen such that the evaluated asymmetry
is sensitive to the anomalous coupling and allows for a comparison 
of cases of different values of the parameter. In what follows, kinematical quantities written 
with a superscript "0" are given in the top quark's rest frame. 
Otherwise, they are given in the laboratory frame. \\

%%%%%%%%%%%%%%%%%%%%%%%%%%%%%%%%%%%%%%%%%%%
\subsection{Lepton polar asymmetry}
%%%%%%%%%%%%%%%%%%%%%%%%%%%%%%%%%%%%%%%%%%%
\begin{figure}[!t]
 \centering
\includegraphics[width=0.46\linewidth]{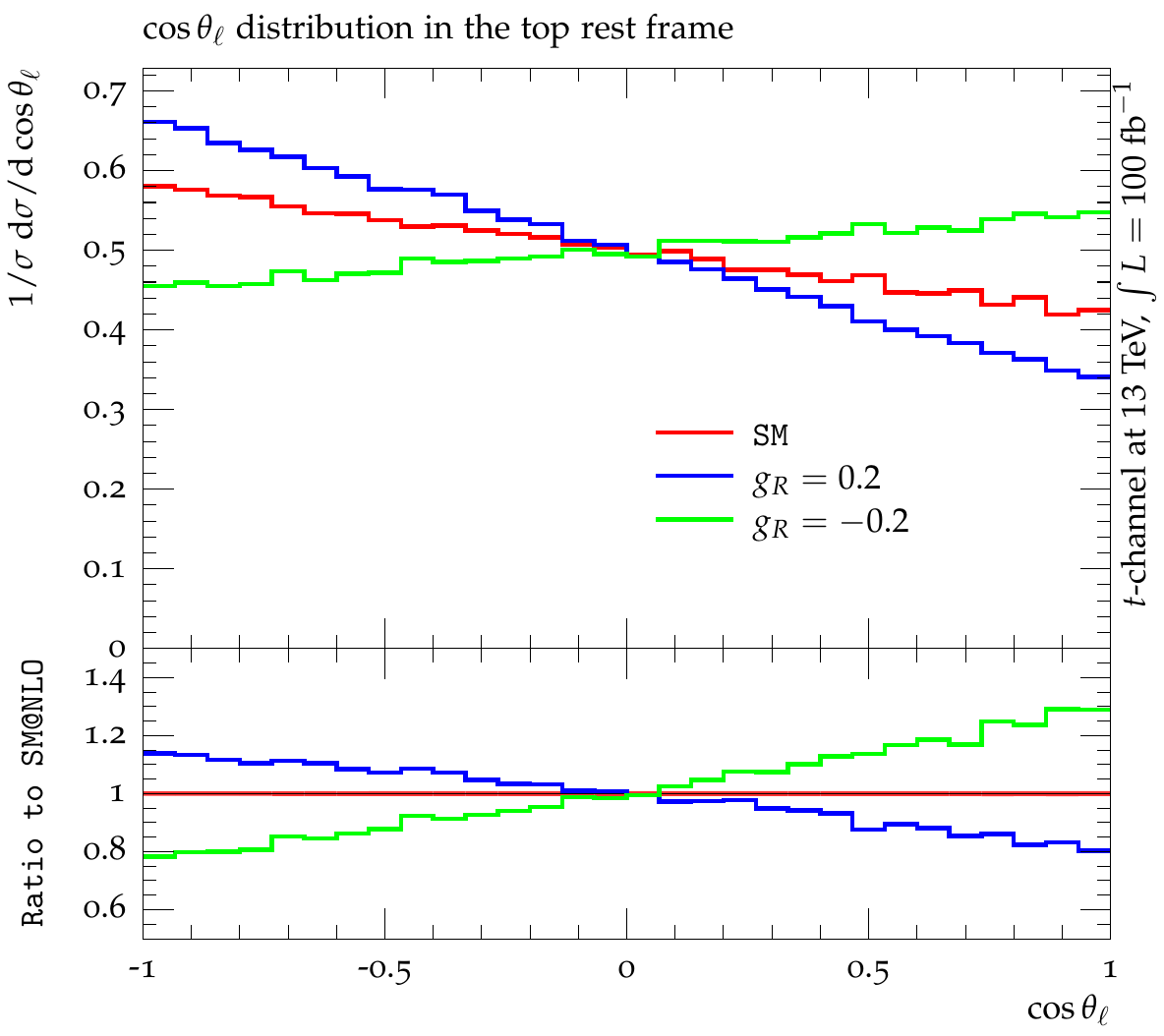}
\hfill
\includegraphics[width=0.46\linewidth]{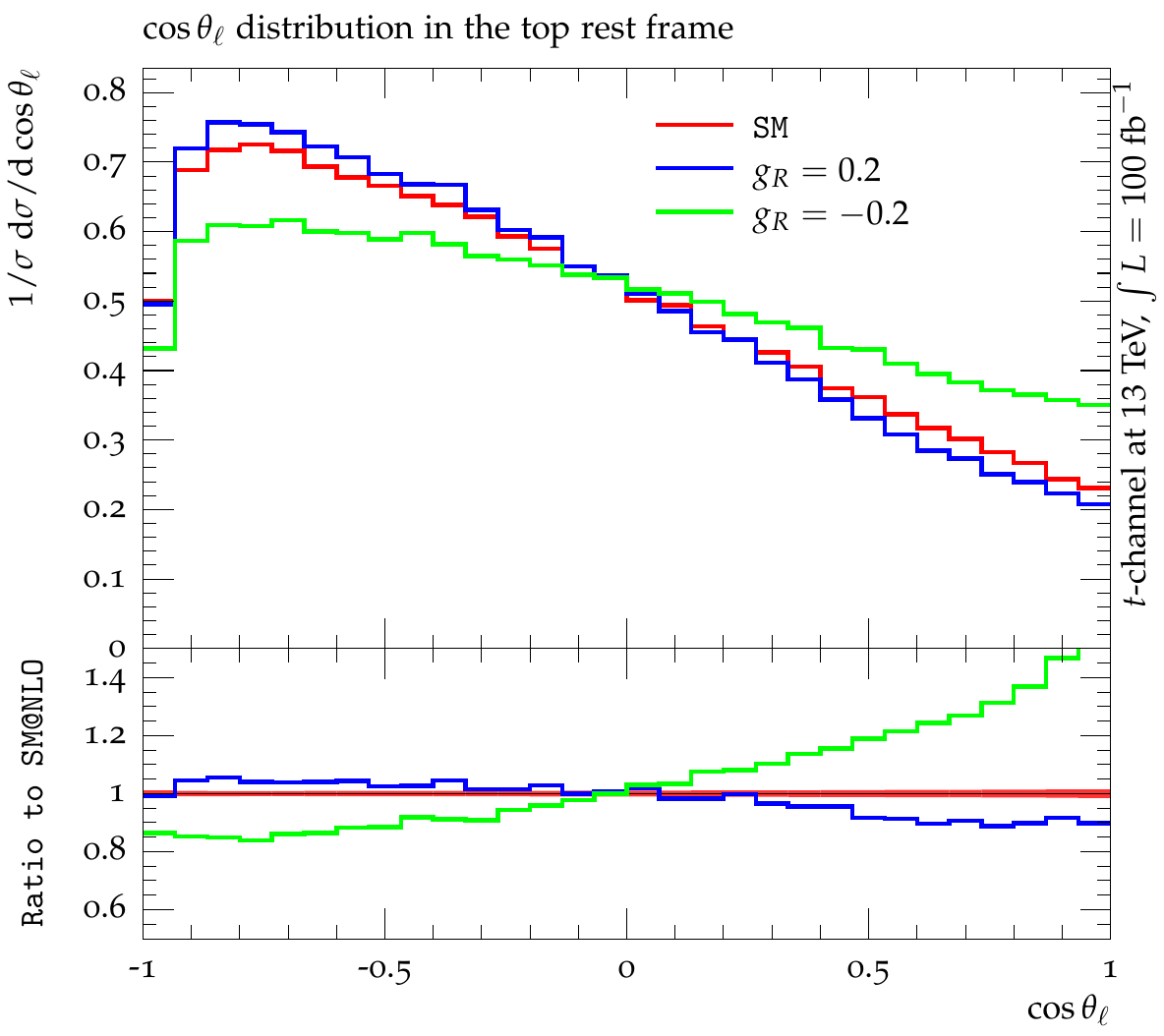}
\caption{$\cos\theta_\ell$ distribution in the top quark rest frame using full partonic
information (left) 
and with the cuts implemented at the particle level (right) at NLO.}
\label{CosThetaRest_NLO}
\end{figure}

The polar angle of the charged lepton, denoted by $\theta_\ell^0$, is defined by
\begin{eqnarray}
 \cos\theta_\ell^0 = \frac{\textbf{p}_\ell \cdot \textbf{p}_t}{|\textbf{p}_\ell| |\textbf{p}_t|},
 \label{pol}
\end{eqnarray}
with $\textbf{p}_\ell$ is the three-momentum of the charged 
lepton in the top quark rest frame and $\textbf{p}_t$ is the top quark 
three-momentum in the laboratory frame. This observable is a good probe of the top quark 
polarisation\footnote{The top quark
direction of flight in the $pp$ center-of-mass 
frame defines the spin quantization 
axis in the helicity basis.}. 
However, given that the presence of anomalous coupling 
changes the chiral structure of the top quark, we expect
that it is a good probe of the anomalous coupling too. This 
can clearly be seen in Fig. \ref{CosThetaRest_NLO} where the 
$\cos\theta_\ell^0$ distribution is plotted in both the SM and for $g_R=\pm0.2$. 
We can see that the effect of the anomalous coupling on the
$\cos\theta_\ell^0$ is important 
and which can even change the slope of the distribution 
for $g_R=-0.2$. Although the presence of cuts 
(right panel of Fig. \ref{CosThetaRest_NLO}) modifies 
the sensitivity of this observable, it can be used for searches 
of the anomalous couplings. However, the measurement of the polar 
distribution is quite challenging since it requires a full 
reconstruction of the top quark momentum. This is hard to be achieved 
due to the presence of missing energy in the semi-leptonic decay of 
the top quark.
Nevertheless, several measurements of the charged lepton 
angle in the helicity basis exist, e.g. in the $t\bar{t}$ system \cite{Aaboud:2016bit}. Hence, it can possibly 
be measured in the $t$-channel process in the future. From Fig. \ref{CosThetaRest_NLO}, we define
the reference point for the $A_{\theta_\ell^0}$ to be $\cos\theta_\ell^0 = 0$.
%%%%%%%%%%%%%%%%%%%%%%%%%%%%%%%%%%%%%%%%%
\subsection{Charged lepton Energy}
%%%%%%%%%%%%%%%%%%%%%%%%%%%%%%%%%%%%%%%%%
In addition to the angular polar distribution in the top
quark's rest frame, two other observables constructed from 
the charged lepton energy are considered here. We define 
a dimensionless variable, $x_\ell$, by:
\begin{eqnarray}
 x_\ell = \frac{2 E_\ell}{m_t},
\label{xL}
 \end{eqnarray}
where $E_\ell$ is the lepton's energy in a given frame and $m_t$ is 
the top quark mass. We consider the energy of the charged lepton in two
different frames; the top quark's rest frame and the $pp$ center-of-mass frame.

\begin{figure}[!t]
 \centering
\includegraphics[width=0.46\linewidth]{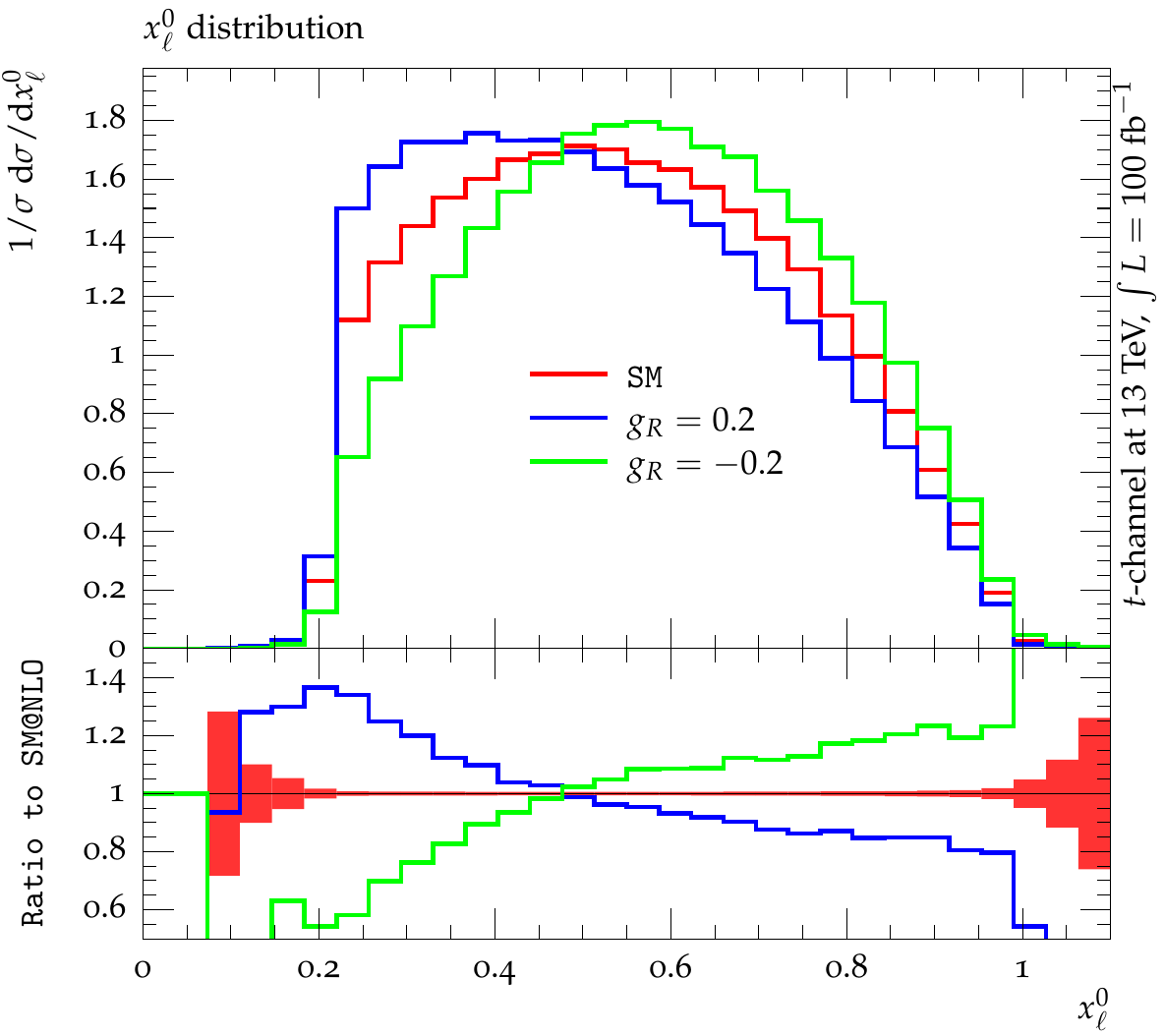}
\hfill
\includegraphics[width=0.46\linewidth]{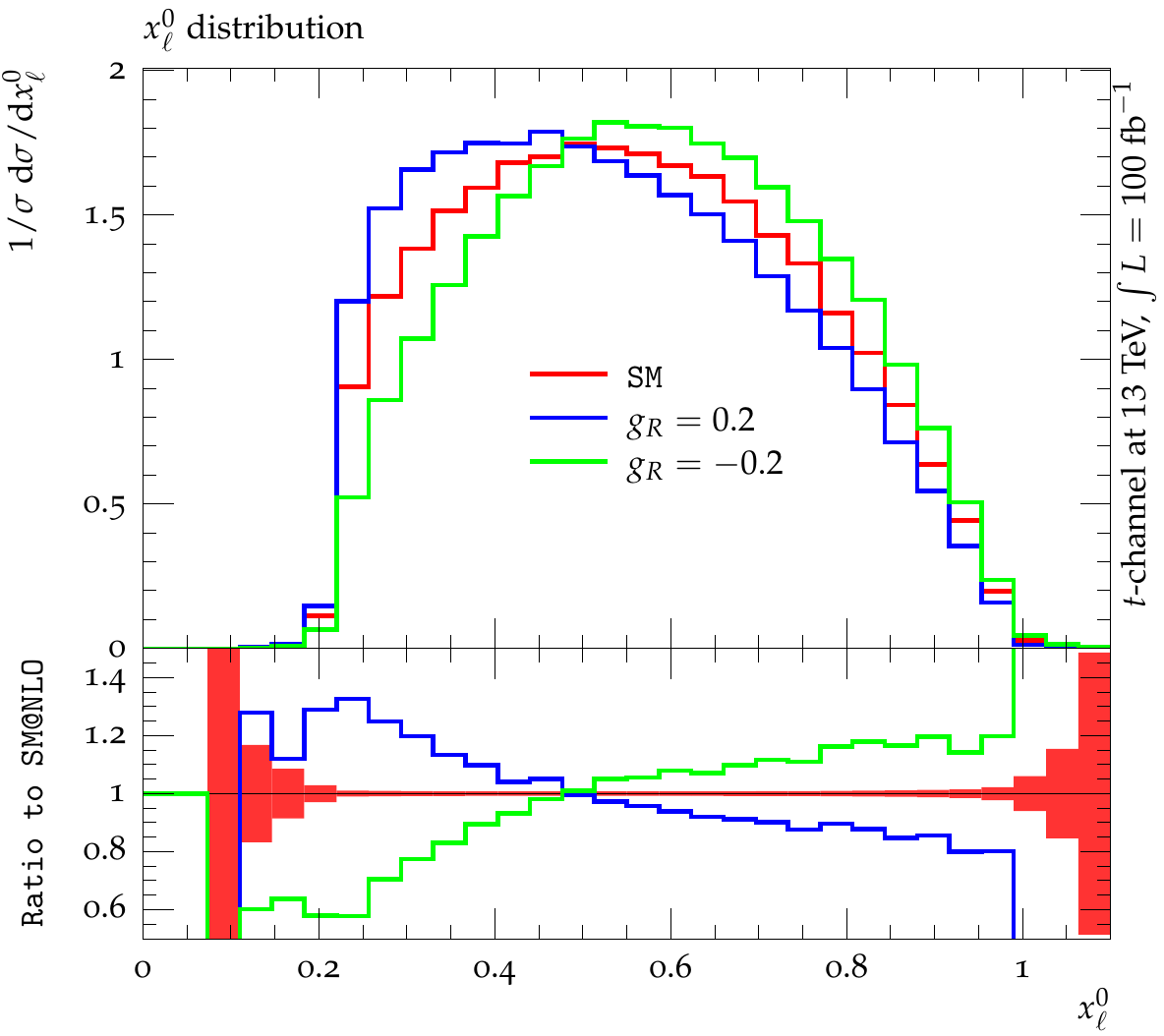}
\caption{$x_\ell$ distribution in the top quark rest frame using full partonic
information (left) 
and with the cuts implemented at the particle level (right) at NLO.}
\label{xLrest_NLO}
\end{figure}

It was shown that, in the top quark's rest frame, 
this asymmetry is a pure probe of the anomalous $Wtb$ coupling regardless
the top quark production mechanism (or in other words top quark 
polarisation) \cite{Prasath:2014mfa}. Hence, at the experimental level, full advantage 
of this observable should be taken by measuring it in several channels. 
However, for its measurement, a reconstruction of the top quark momentum is needed. We
depict the $x_\ell^0$ in the SM and for $g_R=\pm0.2$ in Fig. \ref{xLrest_NLO}. 
The reference point for the corresponding asymmetry, $A_{x_\ell^0}$ is chosen 
to be $x_{\ell^0,c}=0.5$, i.e the value $x_\ell^0$ at the peak position in
the SM \cite{Godbole:2015bda}.

\begin{figure}[!t]
 \centering
\includegraphics[width=0.46\linewidth]{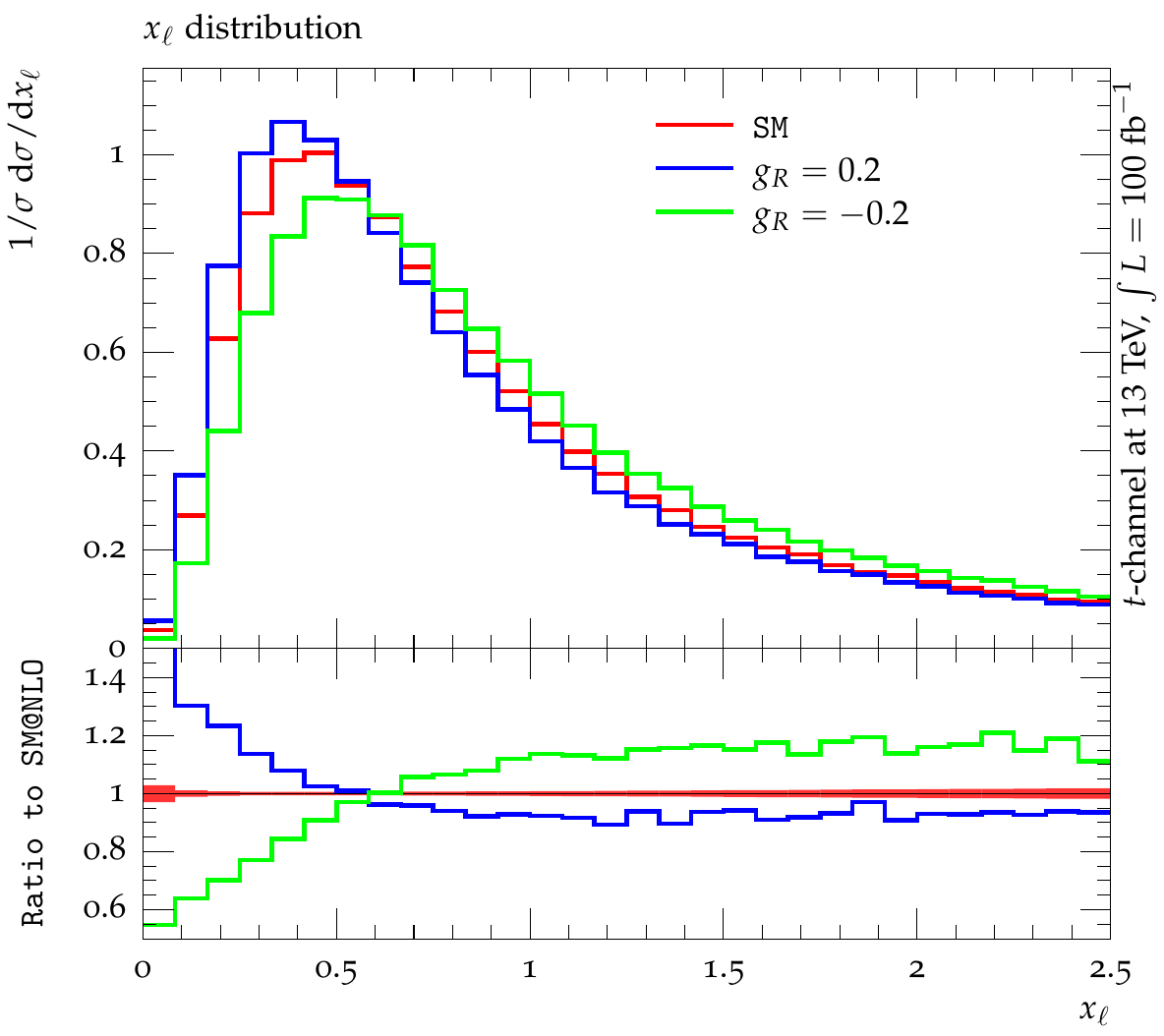}
\hfill
\includegraphics[width=0.46\linewidth]{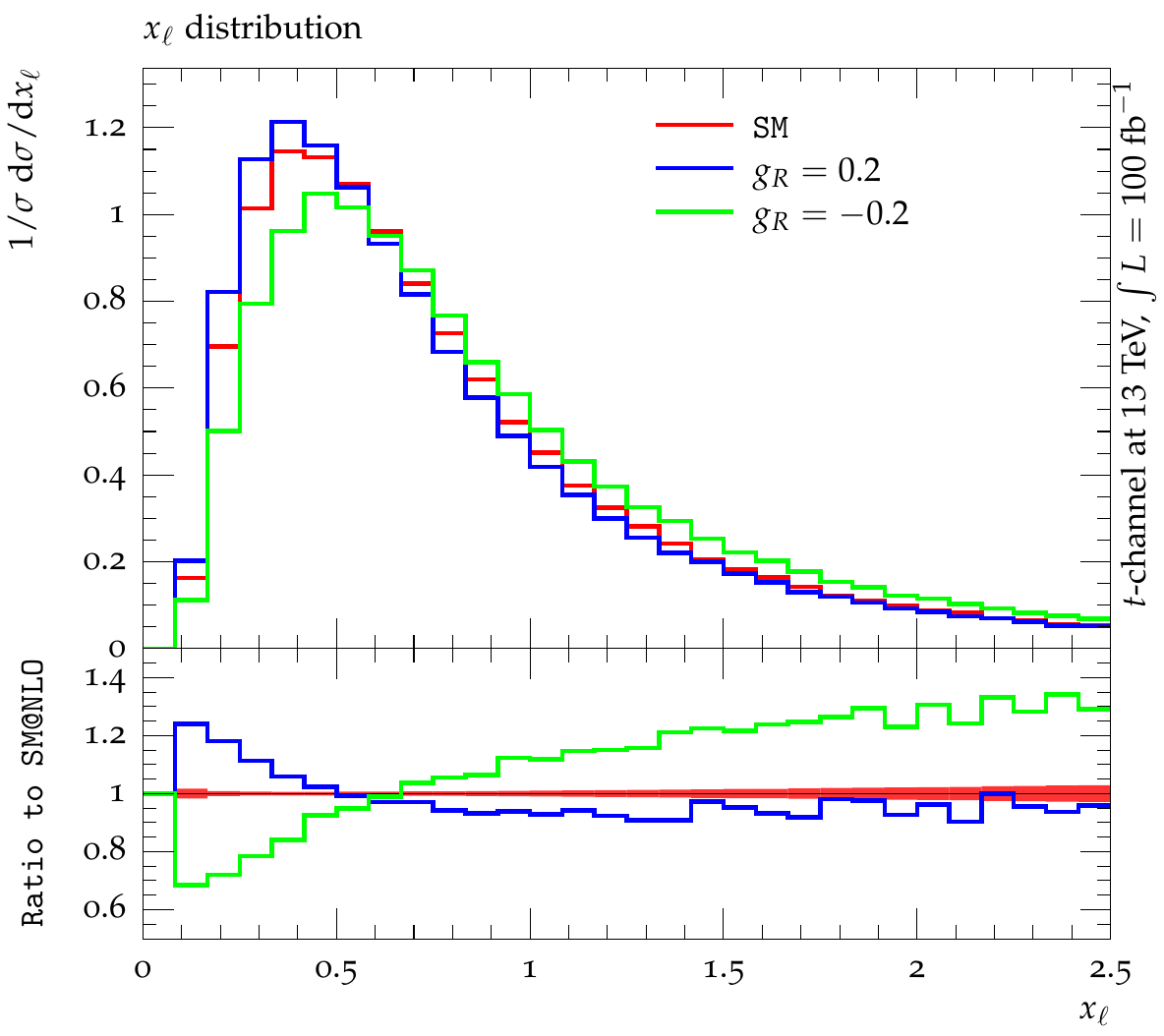}
\caption{$x_\ell$ distribution in the laboratory frame using full partonic
information (left) 
and with the cuts implemented at the particle level (right) at NLO.}
\label{xL_lab_NLO}
\end{figure}

The situation is different for $x_\ell$ in the laboratory frame which is
shown in Fig. \ref{xL_lab_NLO}. This observable is sensitive to both the 
anomalous coupling as well as the top 
polarisation and  
has a high sensitivity to the anomalous
coupling for small values of $x_\ell$ for positive 
values of $g_R$ and in the full range of $x_\ell$ for negative 
values of $g_R$. However, this observable has a lower sensitivity on the anomalous 
coupling  than $x_\ell^0$ due to some cancellations which occur
between anomalous couplings and other kinematical factors \cite{Prasath:2014mfa}. 
Finally, no reconstruction of the top quark
momentum is needed in order to measure this observable. A reference point 
of $x_{\ell,c}=0.6$ is chosen for the evaluation of the corresponding asymmetry \cite{Godbole:2015bda}.

%%%%%%%%%%%%%%%%%%%%%%%%%%%%%%%%%%%%%%%%%%
\subsection{$u$- and $z$-variables}
%%%%%%%%%%%%%%%%%%%%%%%%%%%%%%%%%%%%%%%%%%
\begin{figure}[!t]
 \centering
\includegraphics[width=0.46\linewidth]{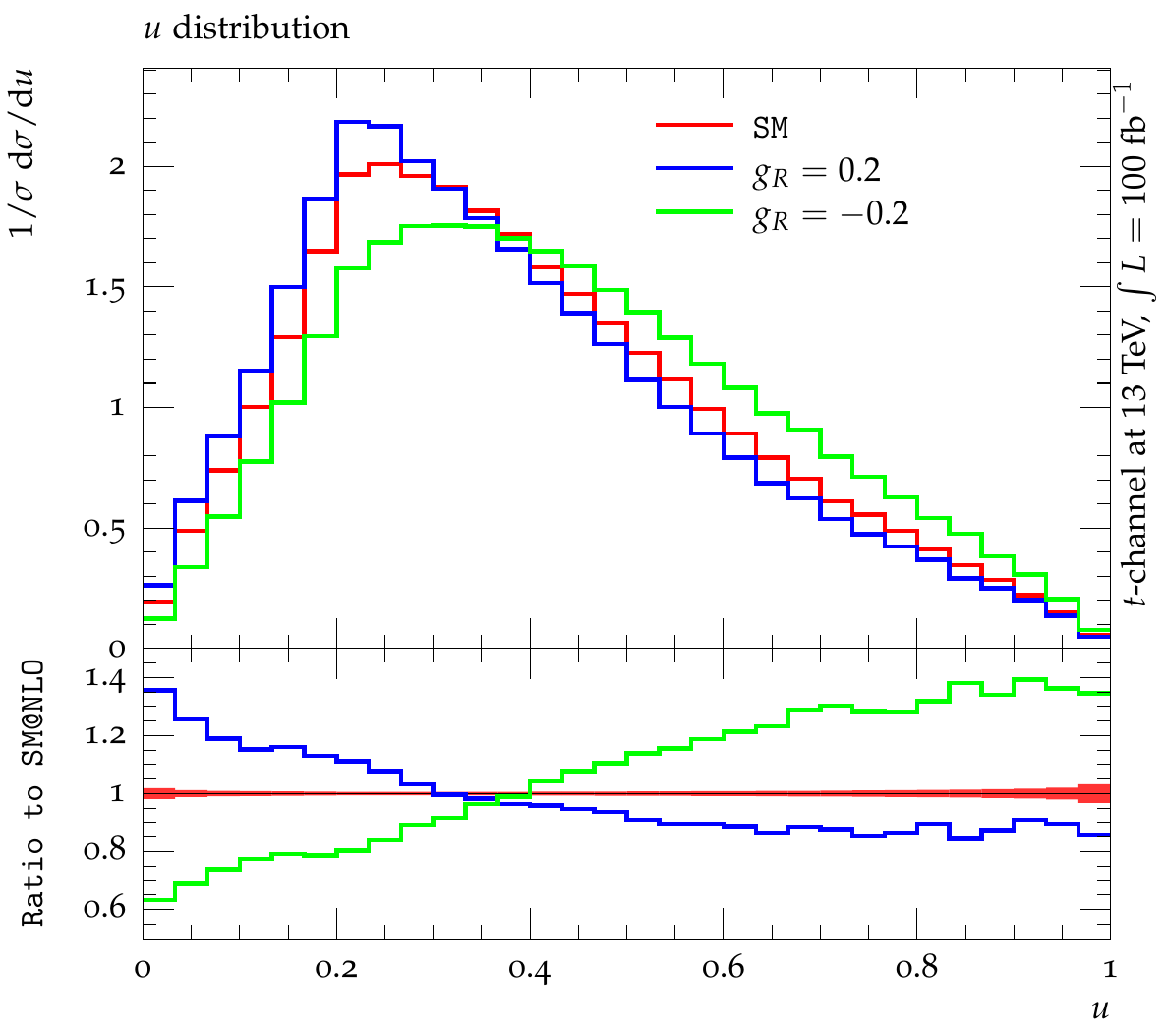}
\hfill
\includegraphics[width=0.46\linewidth]{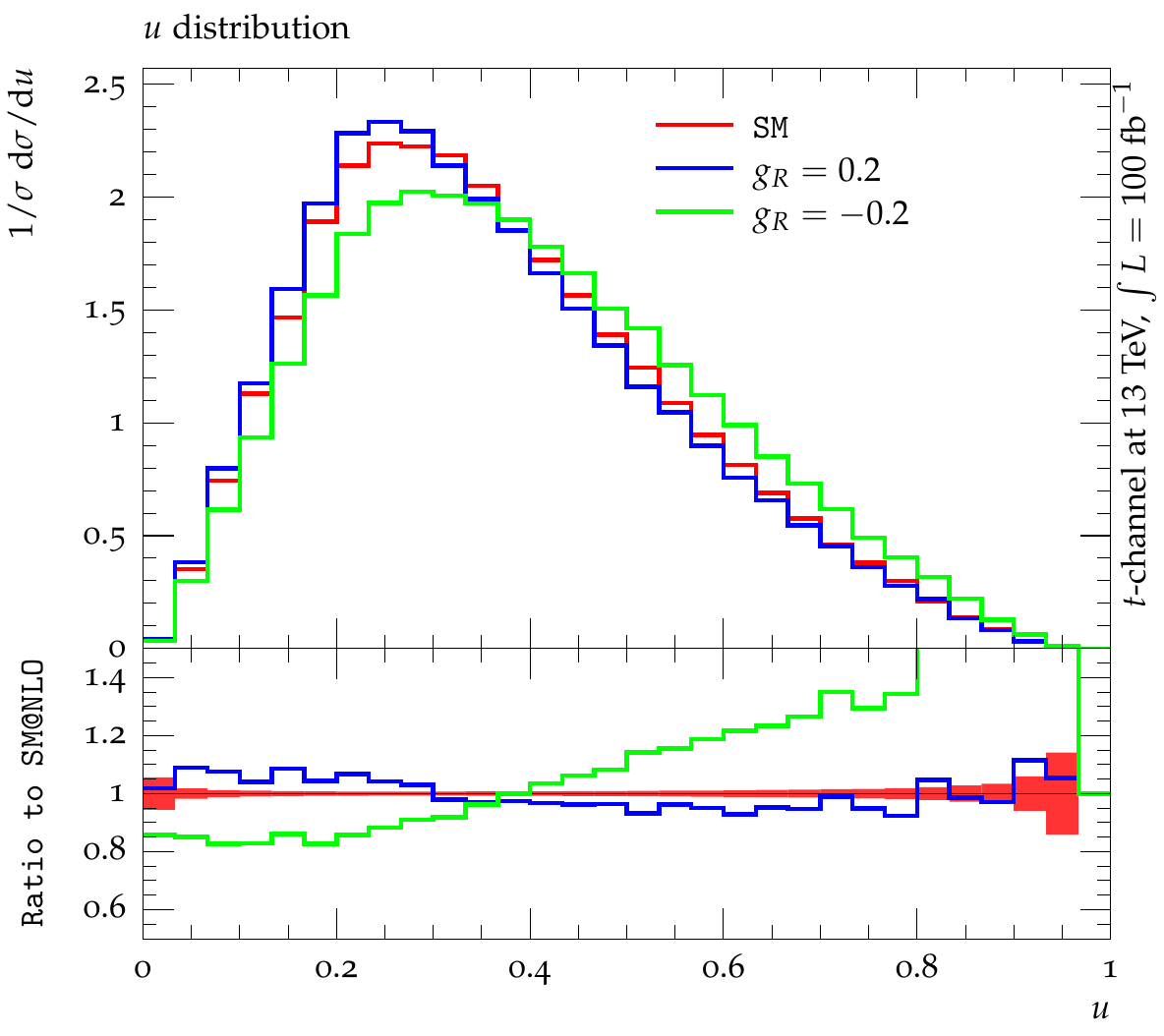}
\caption{$u$ distribution in the laboratory frame using full partonic
information (left) 
and with the cuts implemented at the particle level (right) at NLO.}
\label{u_NLO}
\end{figure}

The last two variables from which two different asymmetries will be evaluated have been 
proposed in \cite{Shelton:2008nq} as a probe of top quark polarisation for highly boosted top 
quarks.
First, a variable $u$ which measures the energy ratio of the charged lepton to the 
total visible energy (of the top quark decay). This variable is defined by :
\begin{eqnarray}
 u = \frac{E_\ell}{E_\ell + E_b},
 \label{uBL}
\end{eqnarray}
where $E_\ell$ and $E_b$ are the lepton and b-quark ($b$-jet) energies 
in the laboratory frame. It was found that the $u$-variable is sensitive 
to both the top quark polarisation and the anomalous $Wtb$ coupling \cite{Prasath:2014mfa}. 
We found that including the anomalous $Wtb$ in the production
improves the sensitivity of the $u$-variable on $g_R$. From experimental point of view, it is possible 
to measure this variable from a simultaneous measurements of the both the charged lepton and $b$-jet energies in
the laboratory frame. This implies that there is no need for reconstructing the top quark momentum. 
We depict the $u$ variable for three different models
at NLO in Fig. \ref{u_NLO}. From this figure, we choose the reference point $u_c = 0.4$ which 
is the intersection point of the three different curves corresponding to the SM
and to $g_R=\pm0.2$.

\begin{figure}[!t]
 \centering
\includegraphics[width=0.46\linewidth]{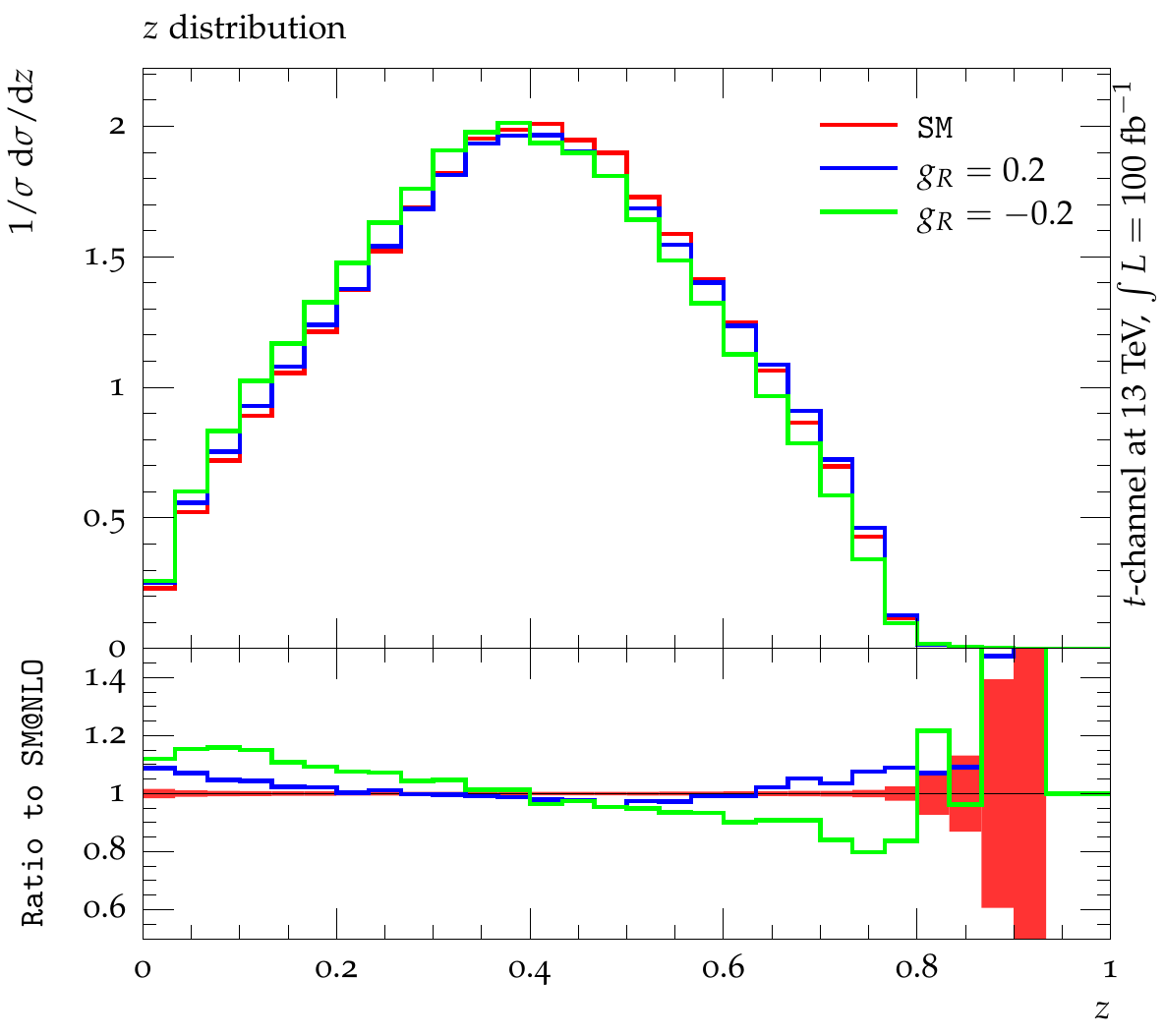}
\hfill
\includegraphics[width=0.46\linewidth]{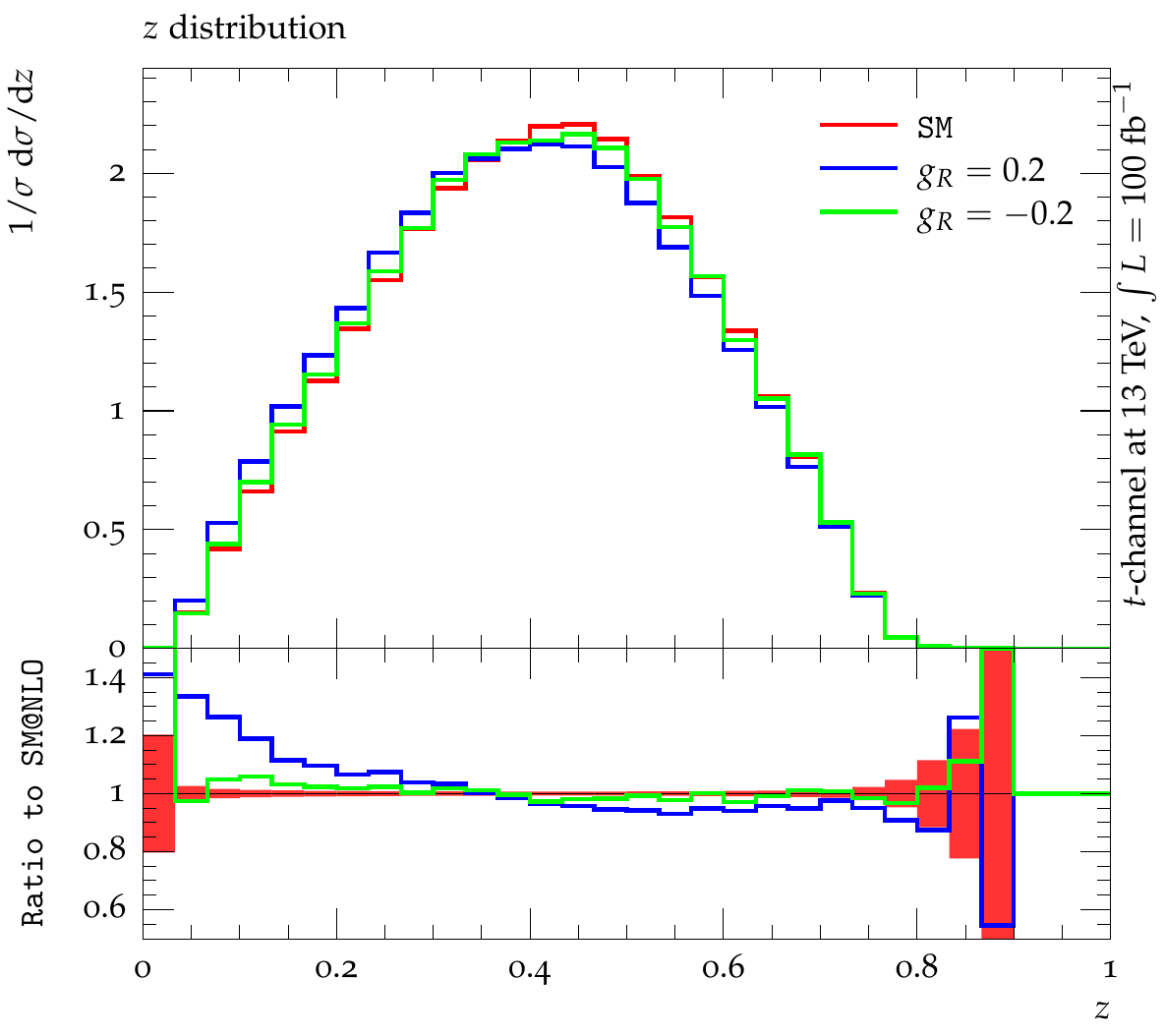}
\caption{$z$ distribution in the laboratory frame using full partonic
information (left) 
and with the cuts implemented at the particle level (right) at NLO.}
\label{z_NLO}
\end{figure}

Finally, the $z$ variable, which measures the fraction of the top quark energy taken by 
the $b$-jet in the laboratory frame, is defined by:
\begin{eqnarray}
 z = \frac{E_b}{E_t},
 \label{zBT}
\end{eqnarray}
where $E_b$ and $E_t$ are the energies of the bottom and 
top quarks respectively in the laboratory frame. In Fig. \ref{z_NLO}, we depict 
the $z$-variable for the SM and $g_R=\pm0.2$. We can see that $z$-variable has a lower
sensitivity on the anomalous coupling. Furthermore, its measurement requires a determination of 
the top quark energy, which by itself depends on the decay mode. The hadronic mode, although has 
a larger background is better for the measurement of the $z$-variable. 
The reference point will be chosen to be $z_c=0.4$. 

\begin{table}[!t]
\begin{center}
 \begin{tabular}{ c | c | c | c }
  \hline \hline
           &   \hspace{1.5cm}  LO \hspace{1.5cm}      &    \hspace{1.5cm}  NLO   \hspace{1.5cm} & \hspace{1cm} NLO/LO \hspace{1cm} \\ \hline
 $A_{\theta_\ell^0}$  &    $0.110^{+0.0000(+0.0\%)}_{-0.01630(-14.5\%)}$  &          $0.104^{+0.0000(+0.0\%)}_{-0.0155(-14.9\%)}$  & $0.94$\\
             &   $0.347^{+0.0000(+0.0\%)}_{-0.0019(-0.5\%)}$      &     $0.297^{+0.0045(+1.5\%)}_{-0.0023(-0.7\%)}$ & $0.85$\\ \hline
 $A_{x_\ell^0}$ &   $-0.139^{+0.0175(+12.5\%)}_{-0.0000(-0.0\%)}$         &          $-0.139^{+0.0010(+0.7\%)}_{-0.0000(-0.0\%)}$  & $1.0$ \\
               &   $-0.178^{+0.0000(+0.0\%)}_{-0.0181(-10.2\%)}$          &          $-0.178^{+0.0025(+1.4\%)}_{-0.0000(-0.0\%)}$ & $1.0$ \\ \hline
$A_{x_\ell}$  &  $-0.117^{+0.0194(+16.5\%)}_{-0.0000(-0.0\%)}$           &          $-0.091^{+0.0000(+0.0\%)}_{-0.0063(-6.9\%)}$ & $1.3$ \\ 
               &  $-0.089^{+0.0000(+0.0\%)}_{-0.0117(-13.1\%)}$           &          $-0.119^{+0.0073(+6.1\%)}_{-0.0039(-3.2\%)}$ & $1.3$ \\ \hline
  $A_u$      &  $0.133^{+0.0000(+0.0\%)}_{-0.0019(-1.4\%)}$                &         $0.131^{+0.0000(+0.0\%)}_{-0.0130(-9.9\%)}$ & $1.0$ \\
            &  $0.271^{+0.0106(+3.9\%)}_{-0.0000(-0.0\%)}$               &          $0.242^{+0.0019(+0.7\%)}_{-0.0038(-1.5\%)}$ & $0.9$ \\ \hline
 $A_z$     &        $-0.090^{+0.0120(+13.3\%)}_{-0.0000(-0.0\%)}$         &          $-0.074^{+0.0000(+0.0\%)}_{-0.0024(-3.0\%)}$ & $0.8$ \\
           &        $-0.013^{+0.0000(+0.0\%)}_{-0.0074(-56.9\%)}$         &          $-0.010^{+0.0085(+85\%)}_{-0.0000(-0.0\%)}$ & $0.8$ \\ \hline \hline                    
 \end{tabular}
\end{center}
\caption{Asymmetries and their theoretical uncertainties in the SM at LO and NLO. The first rows for each asymmetry show the 
partonic level results while the second rows the particle level ones.}
\label{Asym-error}
\end{table}

\section{Results}
\label{sec:results}
%================================

\begin{figure}[!t]
  \centering
  \includegraphics[width=7cm, height=10cm]{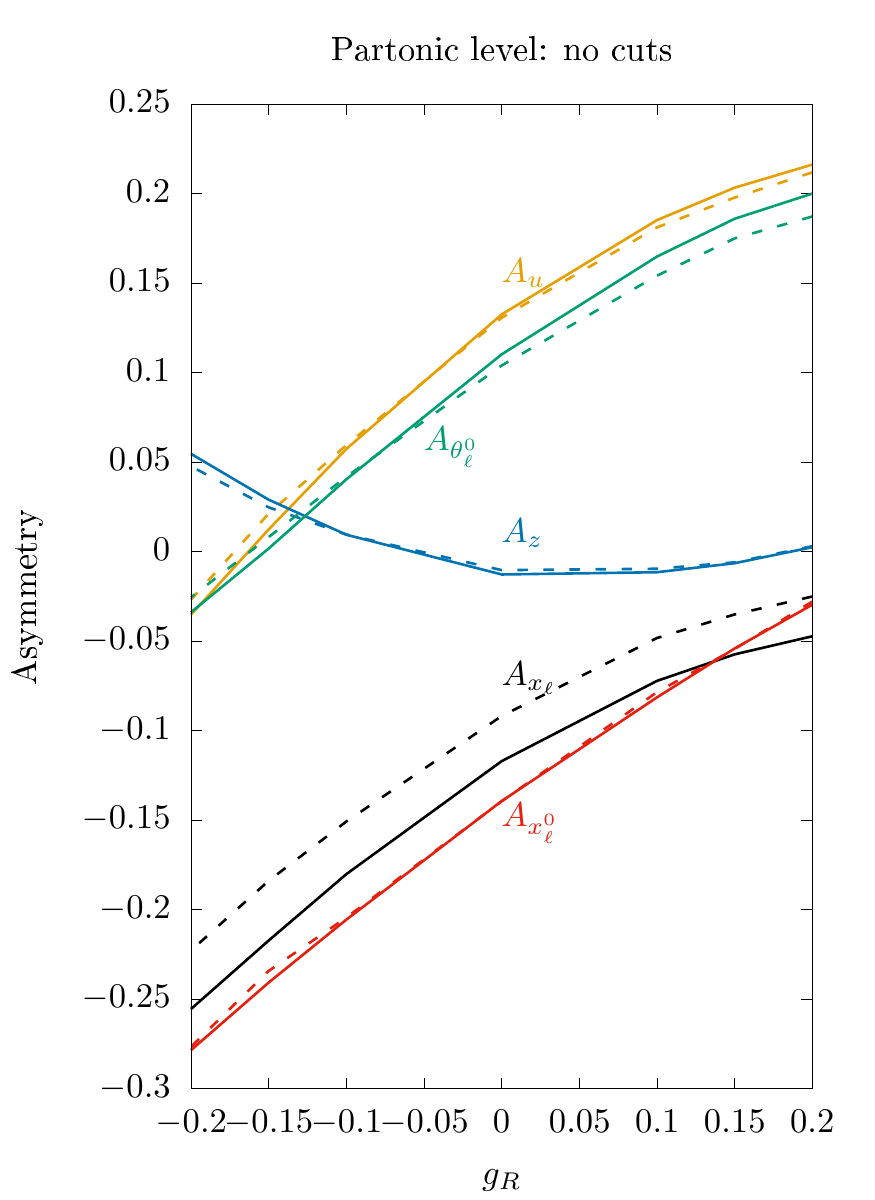}
  \hfill
  \includegraphics[width=7cm, height=10cm]{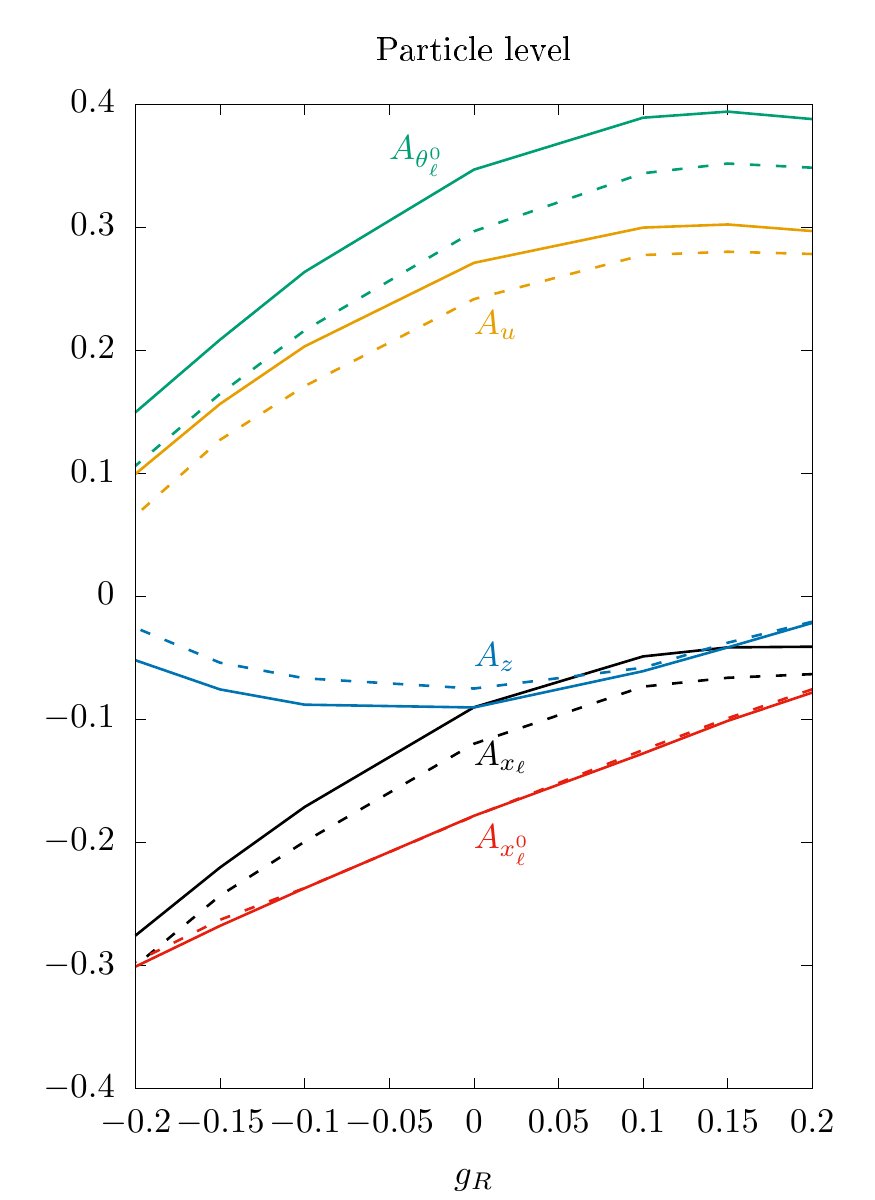}
  \caption{Dependence of different asymmetries as function of the anomalous coupling at
  the parton level without cuts (left 
  panel) and at the particle level with cuts (right panel). The solid (dashed) lines show the 
  dependence at LO (NLO).}
  \label{Asym-plots1}
 \end{figure}

From the different kinematical distributions shown in section \ref{sec:observables}, 
asymmetries are constructed (for more details, see section \ref{sec:appen1}). 
These asymmetries present, except $A_z$, a high sensitivity on the anomalous right tensorial 
coupling as it is depicted in Fig. \ref{Asym-plots1} which is weakened at the particle level.
We can see that there are some differences between the central values of the asymmetries at LO
and NLO of about $6\%$--$30\%$ depending on the particular asymmetry. However, one notices that these differences 
are within the theoretical uncertainties which are depicted in Table \ref{Asym-error} where 
only the effect of scale variations is shown. The effect of radiative corrections on the asymmetries depends
on the particular variable. At the parton level, two asymmetries are perfectly stable against radiative corrections; i.e
$A_{x_\ell^0}$ and $A_u$ while the others can receive corrections of $6\%$ for $A_{\theta_\ell^0}$, $20\%$ for $A_z$
and $30\%$ for $A_{x_\ell}$. On the other hand, the theoretical uncertainties due to scale variations are quite large
except for $A_u$ both at LO and NLO and for $A_{x_\ell^0}$ at NLO. At the particle level, the corrections to the asymmetries
are lower than to the cross sections with again a strong stability against NLO corrections for $A_{x_\ell^0}$ and $A_u$. The theoretical
uncertainties due to scale variations are lower than in the parton level case with one notable exception, i.e $A_z$ which
has very large theoretical uncertainties (see Table \ref{Asym-error}).

We perform a $\chi^2$ exclusion to obtain the limits on the anomalous coupling $g_R$. A 
$\chi^2$ is defined as follows
\begin{eqnarray}
 \chi^2 = \sum_\mathcal{O} \frac{(A_\mathcal{O} - A_{\mathcal{O};\textrm{SM}})^2}
 {\Delta_\mathcal{O}^2},
\end{eqnarray}
where $\mathcal{O}=x_\ell,x_\ell^0,\cos\theta_\ell^0,z \text{ and } u$. 
$\Delta_{\mathcal{O}}$ is the sum, by quadrature, of the statistical and theoretical
uncertainties on the asymmetry $A_\mathcal{O}$ in the SM. The former are defined as
\begin{equation}
 \Delta_{\mathcal{O}}^{\mathrm{stat.}} = \frac{\sqrt{1-A_\mathcal{O}^2}}{\sqrt{N}}.
\end{equation}
where $N$ is the number of events
$$
N = \sigma_{t} \mathrm{BR}(t\to b \ell \nu_\ell) \mathcal{L},
$$
we assume a luminosity of $\mathcal{L} = 100$ fb$^{-1}$. Values of $g_R$ are excluded within 
$1\sigma, 2\sigma \text{ and } 3\sigma$ if the corresponding $\chi^2$ is larger
than $2.3, 5.99$ and $11.8$ respectively.  We show in Table \ref{table-indiv1},
individual limits obtained from the different 
asymmetries at $1\sigma$ both at LO and NLO. We can see that all the asymmetries but $A_z$ give 
strong constraints on the anomalous coupling.
Furthermore, the limits placed on the anomalous coupling from $A_{x_\ell}$ and $A_{\theta_\ell^0}$
are strenghten, at the particle level, by about a factor of $2$ and $7$ respectively. This is due 
to the large theoretical uncertainties on those observables at the partonic
level which are significantly reduced when showers and cuts are implemented. Overall, constraints 
on the anomalous right tensorial coupling are stronger at NLO due to the 
reduction of theoretical and statistical uncertainties. Before discussing the combination of different
asymmetries and its effect on the anomalous coupling, we comment about the experimental uncertainties.
Generally, measurement of the asymmetries
$A_{x_\ell^0}$ and $A_{\theta_\ell^0}$ introduces additional systematic uncertainties since they involve
the reconstruction of the top quark rest frame which is very hard
for single top production. Hence, although the $A_{x_\ell^0}$
asymmertry is resilient to NLO corrections, it might not be very efficient in the determination of
the anomalous coupling. On the other hand, among all the laboratory frame asymmetries, $A_u$ is the most sensitive
and involving less systematics (both theoretical and experimental) 
uncertainties\footnote{However, it is expected that experimental systematic errors drop when
asymmetries are used.}. Combining two asymmetries at one
time improves significantly the limits on $g_R$. We found that combining $A_{x_\ell^0}$ and $A_u$ gives
$[-0.0048,0.0048]$ ($[-0.0021,0.0021]$) at LO (NLO) at the parton level at $1\sigma$. While
at the particle level, those limits are weakened to give $[-0.0261,0.0312]$ ($[-0.0052,0.0053]$) at LO (NLO). \\

These limits can be improved by combining more than one asymmetry at a time. As an example, 
we estimate projected limits using
ten different combinations of the three 
different asymmetrie. We can see
in Tables \ref{Limits.noCuts}-\ref{limits.withCuts} that the limits are improved 
by about one order of magnitude from those obtained using one asymmetry at a time.
In Table \ref{Limits.noCuts}, we show these limits at the partonic level. We can see that, at LO
that six different combinations of the three asymmetries give
$g_R \in [-0.0103,0.0107]$ at $3\sigma$. The situation is different for the NLO case,
where the limits are improved by a factor of $2$ from the combinations $(A_u,A_{x_\ell}, A_{x_\ell^0})$
and $(A_u, A_{x_\ell^0}, A_{\theta_\ell^0})$ yielding $g_R$ in the interval $[-0.0047,0.0048]$. On the other hand,
including showers and cuts do not change much the limits that we obtain but only the combination
of the asymmetries change. In Table \ref{limits.withCuts}, we can see that six asymmetries give the strongest 
limit at LO; i.e $g_R \in [-0.0112, 0.0121]$. While at NLO, we obtain $g_R \in [-0.0099,0.0104]$ using the 
$(A_u, A_{x_\ell^0}, A_{\theta_\ell^0})$ combination. However, as this combination involves the $A_{\theta_\ell^0}$ 
asymmetry which requires full reconstruction of the top quark direction of motion. Hence, that either 
the $(A_u,A_{x_\ell}, A_z)$ or $(A_u,A_{x_\ell}, A_{x_\ell^0})$ will do a better 
job in pinning down the anomalous coupling even the obtained limits are milder
than $(A_u, A_{x_\ell^0}, A_{\theta_\ell^0})$.  
 
\begin{table}[!t]
 \begin{center}
 \begin{tabular}{ c | c | c }
 \hline \hline
  Asymmetry   &  \hspace{0.5cm} \texttt{LO} \hspace{0.5cm} & \hspace{0.5cm} \texttt{NLO} \hspace{0.5cm} \\ \hline 
    $A_{\theta_\ell^0}$ & $[-0.0380, 0.0440]$ & $[-0.0480, 0.0541]$ \\ 
                      & $[-0.0052, 0.0053]$ & $[-0.0082, 0.0086]$ \\ \hline
    $A_{x_\ell^0}$ & $[-0.0417, 0.0446]$   &   $[-0.0026, 0.0026]$ \\
                 & $[-0.0486, 0.0542]$   &   $[-0.0069, 0.0070]$ \\ \hline
     $A_{x_\ell}$   &  $[-0.0500, 0.0605]$    &   $[-0.0193, 0.0204]$ \\ 
                 &  $[-0.0261, 0.0321]$    &   $[-0.0122, 0.0131]$ \\ \hline
    $A_u$       &   $[-0.0046, 0.0047]$    &   $[-0.0036, 0.0036]$ \\ 
              &   $[-0.0302, 0.0402]$    &   $[-0.0080, 0.0085]$ \\ \hline
     $A_z$       &   $[-0.0702, 0.1912]$    &   $[-0.0738, 0.1987]$ \\
              &   $[-0.1756, 0.0876]$    &   $[-0.0702, 0.0490]$ \\ \hline \hline
 \end{tabular}-
\end{center}
\caption{Individual expected limits at $1\sigma$ on the anomalous coupling $g_R$ at LO
and NLO using full partonic information (first rows) and at the particle level (second rows).}
\label{table-indiv1}
\end{table}

\begin{table}[!t]
 \begin{center}
  \begin{tabular}{ c | c | c | c }
  \hline \hline
   Combination             & \hspace{0.5cm}  $1\sigma$   \hspace{0.5cm}      &  \hspace{0.5cm} $2\sigma$  \hspace{0.5cm}  & 
   \hspace{0.5cm} $3\sigma$ \hspace{0.5cm} \\ \hline
   $(A_u,A_z,A_{x_\ell})$  &  \hspace{0.5cm} $[-0.0046,0.0047]$ \hspace{0.5cm} & \hspace{0.5cm}  $[-0.0074, 0.0076]$ \hspace{0.5cm} 
   & \hspace{0.5cm} $[-0.0103, 0.0107]$  \hspace{0.5cm}   \\ 
   & \hspace{0.5cm} $[-0.0035,0.0036]$ \hspace{0.5cm} & \hspace{0.5cm}  $[-0.0057, 0.0058]$ \hspace{0.5cm} 
   & \hspace{0.5cm} $[-0.0079, 0.0082]$  \hspace{0.5cm}   \\ \hline
   $(A_u,A_{x_\ell}, A_{x_\ell^0})$ &  $[-0.0046, 0.0046]$ & $[-0.0074, 0.0076]$ & $[-0.0103, 0.0107]$ \\
                                    &  $[-0.0021, 0.0021]$ & $[-0.0034, 0.0034]$ & $[-0.0047, 0.0048]$ \\ \hline
   $(A_{x_\ell}, A_{x_\ell^0}, A_{\theta_\ell^0})$ & $[-0.0250, 0.0266]$ & $[-0.0397, 0.0438]$ & $[-0.0547, 0.0629]$ \\ 
    & $[-0.0026, 0.0026]$ & $[-0.0042, 0.0042]$ & $[-0.0058, 0.0059]$ \\ \hline 
   $(A_u,A_z, A_{\theta_\ell^0})$ & $[-0.0046, 0.0046]$ & $[-0.0074, 0.0076]$ & $[-0.0103, 0.0107]$ \\ 
    & $[-0.0036, 0.0036]$ & $[-0.0058, 0.0059]$ & $[-0.0080, 0.0083]$ \\ \hline
   $(A_u,A_z,A_{x_\ell^0})$ & $[-0.0046, 0.0046]$ & $[-0.0074, 0.0076]$ & $[-0.0103, 0.0107]$ \\
     & $[-0.0021, 0.0021]$ & $[-0.0034, 0.0034]$ & $[-0.0048, 0.0048]$ \\ \hline 
   $(A_u,A_{x_\ell}, A_{\theta_\ell^0})$ & $[-0.0046, 0.0047]$ & $[-0.0074, 0.0076]$ & $[-0.0103, 0.0107]$ \\ 
     & $[-0.0035, 0.0036]$ & $[-0.0056, 0.0058]$ & $[-0.0079, 0.0082]$ \\ \hline
   $(A_u, A_{x_\ell^0}, A_{\theta_\ell^0})$ &  $[-0.0046, 0.0046]$ & $[-0.0074, 0.0076]$ & $[-0.0103, 0.0107]$ \\ 
     &  $[-0.0021, 0.0021]$ & $[-0.0034, 0.0034]$ & $[-0.0047, 0.0048]$ \\ \hline
   $(A_z,A_{x_\ell}, A_{\theta_\ell^0})$ & $[-0.0294, 0.0335]$ & $[-0.0458, 0.0562]$ & $[-0.0622, 0.0822]$ \\
     & $[-0.0178, 0.0189]$ & $[-0.0283, 0.0312]$ & $[-0.0393, 0.0449]$ \\ \hline
   $(A_z,A_{x_\ell^0}, A_{\theta_\ell^0})$ & $[-0.0274, 0.0299]$ & $[-0.0430, 0.0496]$ & $[-0.0588, 0.0713]$ \\
     & $[-0.0026, 0.0026]$ & $[-0.0042, 0.0042]$ & $[-0.0059, 0.0059]$ \\ \hline
   $(A_z,A_{x_\ell}, A_{x_\ell^0})$ & $[-0.0309, 0.0348]$ & $[-0.0482, 0.0581]$ & $[-0.0654, 0.0840]$ \\ 
   & $[-0.0026, 0.0026]$ & $[-0.0042, 0.0042]$ & $[-0.0059, 0.0059]$ \\ \hline \hline
  \end{tabular}
 \end{center}
 \caption{Limits on the anomalous coupling $g_R$ at $1\sigma$, $2\sigma$ and $3\sigma$
 at the partonic level without cuts. The first row for each combination represents the 
 interval at LO while the second row represents the NLO exclusion.}
 \label{Limits.noCuts}
\end{table}

\begin{table}[!t]
 \begin{center}
  \begin{tabular}{ c | c | c | c }
  \hline \hline
   Combination             & \hspace{0.5cm}  $1\sigma$   \hspace{0.5cm}      &  \hspace{0.5cm} $2\sigma$  \hspace{0.5cm}  & 
   \hspace{0.5cm} $3\sigma$ \hspace{0.5cm} \\ \hline
   $(A_u,A_z,A_{x_\ell})$  &  \hspace{0.5cm} $[-0.0202,0.0233]$ \hspace{0.5cm} & \hspace{0.5cm}  $[-0.0314, 0.0403]$ \hspace{0.5cm} 
   & \hspace{0.5cm} $[-0.0426, 0.0616]$  \hspace{0.5cm} \\    
   &  \hspace{0.5cm} $[-0.0067,0.0070]$ \hspace{0.5cm} & \hspace{0.5cm}  $[-0.0108, 0.0115]$ \hspace{0.5cm} 
   & \hspace{0.5cm} $[-0.0149, 0.0164]$  \hspace{0.5cm}   \\ \hline  
   $(A_u,A_{x_\ell}, A_{x_\ell^0})$ &  $[-0.0189, 0.0215]$ & $[-0.0294, 0.0368]$ & $[-0.0410, 0.0555]$ \\ 
   &  $[-0.0048, 0.0049]$ & $[-0.0078, 0.0080]$ & $[-0.0108, 0.0113]$ \\ \hline
   $(A_{x_\ell}, A_{x_\ell^0}, A_{\theta_\ell^0})$ & $[-0.0050, 0.0052]$ & $[-0.0081, 0.0085]$ & $[-0.0112, 0.0121]$ \\ 
   & $[-0.0049, 0.0049]$ & $[-0.0078, 0.0081]$ & $[-0.0110, 0.0114]$ \\ \hline
   $(A_u,A_z, A_{\theta_\ell^0})$ & $[-0.0051, 0.0053]$ & $[-0.0081, 0.0086]$ & $[-0.0113, 0.0122]$ \\ 
   & $[-0.0058, 0.0059]$ & $[-0.0092, 0.0097]$ & $[-0.0128, 0.0138]$ \\ \hline
   $(A_u,A_z,A_{x_\ell^0})$ & $[-0.0259, 0.0300]$ & $[-0.0461, 0.0515]$ & $[-0.0541, 0.0768]$ \\ 
   & $[-0.0053, 0.0054]$ & $[-0.0085, 0.0087]$ & $[-0.0118, 0.0123]$ \\ \hline
   $(A_u,A_{x_\ell}, A_{\theta_\ell^0})$ & $[-0.0050, 0.0052]$  & $[-0.0080, 0.0084]$ & $[-0.0114, 0.0120]$ \\ 
   & $[-0.0052, 0.0054]$  & $[-0.0084, 0.0088]$ & $[-0.0117, 0.0125]$ \\ \hline
   $(A_u, A_{x_\ell^0}, A_{\theta_\ell^0})$ & $[-0.0051, 0.0052]$ & $[-0.0081, 0.0085]$ & $[-0.0113, 0.0121]$ \\ 
   & $[-0.0044, 0.0045]$ & $[-0.0071, 0.0073]$ & $[-0.0099, 0.0104]$ \\ \hline
   $(A_z,A_{x_\ell}, A_{\theta_\ell^0})$ & $[-0.0051,0.0052]$ & $[-0.0081, 0.0086]$ & $[-0.0112, 0.0121]$ \\ 
   & $[-0.0068,0.0071]$ & $[-0.0109, 0.0116]$ & $[-0.0152, 0.0165]$ \\ \hline
   $(A_z,A_{x_\ell^0}, A_{\theta_\ell^0})$ & $[-0.0051, 0.0052]$ & $[-0.0082, 0.0086]$ & $[-0.0114, 0.0123]$ \\ 
   & $[-0.0053, 0.0054]$ & $[-0.0085, 0.0087]$ & $[-0.0119, 0.0123]$ \\ \hline
   $(A_z,A_{x_\ell}, A_{x_\ell^0})$ & $[-0.0231, 0.0264]$ & $[-0.0363, 0.0447]$ & $[-0.0495, 0.0661]$ \\ 
   & $[-0.0060, 0.0061]$ & $[-0.0097, 0.0100]$ & $[-0.0136, 0.0140]$ \\ \hline \hline
  \end{tabular}
 \end{center}
\caption{Same as Table \ref{Limits.noCuts} but at the particle level.}
\label{limits.withCuts}
\end{table} 

Now, let us compare our findings with the other results 
in the literature which used different observables in different channels; $W$-boson angular observables, 
single top production cross sections at hadron colliders...etc. In ref. \cite{Boos:1999dd}, limits were obtained from single top production
at both the Tevatron and the LHC, they obtained $-0.12 < g_R < 0.13$ for the LHC
and $-0.24 < g_R < 0.25$ for the Tevatron. In their 
simulation, they considered $pp,p\bar{p}\to Wb\bar{b}$ and 
$pp,p\bar{p}\to Wb\bar{b}+\textrm{ jet}$ processes 
at tree level and included 
$V_L, g_L \text{ and } g_R$ anomalous couplings using both background contribution
as well signal contribution from top quark decays (i.e $s$-channel process). 
In \cite{Hubaut:2005er}, a detailed study of ATLAS sensitivity to 
the top quark and $W$-boson polarisation in $t\bar{t}$ production using both 
semi-leptonic and dileptonic channels was carried out. This analysis was translated 
to limits on the anomalous $Wtb$ couplings, top quark decay into charged Higgs boson 
and to constraints on resonances. Using $W$ boson polarisation, and assuming 
the presence of all the anomalous couplings with $V_L=V_{tb}$,  they got limits 
on $g_R$, i.e $g_R \in [-0.065, 0.070]$ at $3\sigma$. Limits on $g_R$ were 
obtained by the authors of \cite{AguilarSaavedra:2006fy} using
helicity fractions, some angular and energetic asymmetries and spin-spin correlations 
observables in $t\bar{t}$ production at the LHC. They obtained $-0.019 < g_R < 0.018$ at the 
$1\sigma$ level by using the $A_+$ asymmetry defined as 
$$
A_+ = 3 \beta [F_0 + (1 + \beta) F_R]
$$
where $F_{0,R}$ are $W$-boson helicity fractions and $\beta=2^{1/3}-1$. In their work, 
they assumed the presence of all anomalous $Wtb$ couplings. In \cite{AguilarSaavedra:2007rs}, 
sensitivity of the ATLAS experiment on the 
anomalous $Wtb$ couplings was studied. The study concerned $t\bar{t}$ production 
with semi-leptonic final state. In \cite{AguilarSaavedra:2007rs}, $W$-boson helicity
fractions, ratios of helicity fractions and new angular asymmetries were studied. 
A systematic study of the different background 
contributions was carried including detector effects and particle reconstruction 
effeciencies. All 
the anomalous couplings were included and individual limits 
on $g_R$ were obtained by setting $g_L=V_R=0$ and $V_L=V_{tb}$. The stringent bound on 
$g_R$ was obtained from the $A_-$ asymmetry, i.e $g_R \in [-0.0166,0.0282]$. By setting 
two couplings to be non-zero at a time and combining four measurements, they got 
the strongest constraint on $g_R$, i.e $[-0.0108,0.0175]$ which obtained in combination 
with $g_L$ and considerably improves the limit from $A_+$ alone. In \cite{Najafabadi:2008pb}, 
limits on $g_{L,R}$ were obtained using 
cross section of 
single top production through $tW$ channel at the partonic level assuming 
$V_L=V_{tb}$ and $V_R=0$. They obtained $g_R \in [-0.105,0.041]$ assuming 
a $25\%$ systematic uncertainties. In reference \cite{AguilarSaavedra:2008gt}, limits on anomalous $Wtb$
couplings were obtained using single top production 
at the LHC. On the one hand, they obtained limits by combining single top production
cross section through $t$-, $s$- and $tW$ channels with the ratio $R$ defined by 
\begin{eqnarray*}
 R = \frac{\sigma(pp\to\bar{t}+j)}{\sigma(pp\to t+j)}.
\end{eqnarray*}
They found $g_R\in[-0.10, 0.14]$ assuming $V_R=0, V_L=V_{tb}$ and $g_L\neq0$. Then, 
they included top quark decay observables such as $W$-boson helicity fractions and their 
ratios. The obtained limit is $-0.012 < g_R < 0.024$ which is about one order of magnitude
better than those obtained from cross section measurements alone. In \cite{AguilarSaavedra:2010nx}, 
limits on anomalous $Wtb$ couplings were obtained using 
new proposed observables which consist of angular distributions which probe the $W$ boson polarisation
and assuming that only one coupling is non-zero at a time or 
they are either purely real or purely imaginary. They got the following limits at $3\sigma$
\begin{eqnarray*}
|\RE(g_R)| > 0.056 \textrm{  from measurement of } A_+, \\
|\IM(g_R)| > 0.115 \textrm{  from measurement of } A_{FB}^N,
\end{eqnarray*}
where $A_{FB}^N$ is a proposed asymmetry which vanishes for real 
anomalous couplings. \\

The authors of \cite{Rindani:2011pk}, derived limits on the anomalous coupling $g_R$ in $tW^-$ production
at the LHC using top quark polarisation, charged lepton energy distribution and azimuthal asymmetry
at $7$ and $14$ TeV, they obtained the limit $-0.010 < \text{Re}(g_R) < 0.015$ at $1\sigma$
from the combination of three asymmetries assuming 
$V_L=1 \text{ and } V_R=g_L=0$. The authors of \cite{AguilarSaavedra:2011ct} obtained limits on the anomalous
$Wtb$ couplings from the ATLAS and CMS measurements of single top quark production
cross section through $t$-channel and top quark decay observables at $7$ TeV at 
$95\%$ \textbf{CL}. Different limits on $V_L,V_R, g_L \text{ and } g_R$ were obtained
by different combinations of the observables. In \cite{Dutta:2013mva}, limits 
were obtained by considering top quark production
at a future Large Hadron electron Collider (LHeC). They proved that the sensitivity
on the different anomalous couplings can be significantly improved in this new collider environment using
several angular observables. Using $tW^-$ production at the LHC at 
$\sqrt{s}=7\oplus14$ TeV, the authors of \cite{Rindani:2015vya} obtained 
limits on the tensorial right coupling and the anomalous top-gluon coupling by combining 
the azimuthal asymmetry, top quark polarisation and energy asymmetries of the $b$-quark
and the charged lepton. They got $g_R \in [-0.03, 0.08]$ at the 
$1\sigma$ level at the parton level.

%%%%%%%%%%%%%%%%%%%%%%%%%%%%%%%%%%%%%%%%%%%
\section{Conclusions}
\label{sec:conclusions}
%%%%%%%%%%%%%%%%%%%%%%%%%%%%%%%%%%%%%%%%%%%
In this work, we studied the senstivity of asymmetries constructed from 
energy and angular distributions of the top quark's decay products on the 
anomalous right tensorial coupling in single top production through $t$-channel at 
the LHC at $\sqrt{s}=13$ TeV and with an effective luminosity of $100$ fb$^{-1}$. 
We included for the first time the contribution of the anomalous coupling in the production with  
NLO effects. The study was carried both at the parton level and 
at the particle level with some loose cuts applied on different
kinematical variables. We found that asymmetries in the laboratory frame
are more suitable to constrain anomalous couplings. However, 
it is worth to investigate the potential of rest frame observables in the search 
of anomalous couplings although they need a reconstruction of top quark 
rest frame. Moreover, these observables present 
some resilience, within theoretical uncertainties,
to next-to-leading order corrections. Furthermore, We found that combination of different 
asymmetries at one time gives even stronger limits. With their important sensitivity on the 
anomalous coupling, these observables 
are competitive with $W$-boson helicity fractions and other related observables. 
Hence, taking 
into account these observables for future experimental searches seems to be indispensable. 

\section*{Acknowledgements}
The author would like to thank A. Arhrib, F. Boudjema and 
R. M. Godbole for their encouragements to do the study, and their critical comments about 
the manuscript. The author would like to thank O. Mattelaer, E. Re, P. Skands and M. Zaro for
useful discussions about MC event generation and F. Cornet and S. Nasri for their comments.
This work was supported by the Moroccan Ministry of Higher
Education and Scientific Research MESRSFC and  CNRST: 
"Projet dans les domaines prioritaires
de la recherche scientifique et du d\'eveloppement technologique": PPR/2015/6, by Sandwich 
Educational and Training Program (ICTP), by the 
GDRI-P2IM Maroc-France (LAPTh), by ENIGMASS and by
Shanghai Pujiang Program. The author would like thank ICTP (Trieste) and LPSC (Grenoble) 
for providing computational facilities.
\appendix
\section{Interpolations}
\label{sec:appen1}
\begin{table}[!h]
  \begin{center}
  \begin{tabular}{ c | c | c | c | c | c | c | c }
  \hline \hline
  $A_\mathcal{O}$     &  $\zeta_0$ & $\zeta_1$ & $\zeta_2$ & $\zeta_3$ & $\zeta_4$ & $\zeta_5$ & $\zeta_6$ \\ \hline \hline
  $A_{\theta_\ell^0}$ & $0.110$ & $0.620$ & $-0.792$ & $0.526$ & $2.850$ & $-35.518$ & $-0.962$  \\
                      & $0.104$ & $0.558$ & $-0.673$ & $0.630$ & $9.076$ & $-32.390$ & $-167.54$  \\ \hline \hline
  $A_{x_\ell^0}$ & $-0.139$ & $0.618$ & $-0.493$ & $0.299$ & $9.457$ & $-4.994$ & $-159.285$ \\
                 & $-0.139$ & $0.680$ & $-0.229$ & $-6.307$ & $4.152$ & $121.203$ & $-161.842$ \\ \hline \hline
  $A_{x_\ell}$  & $-0.117$ & $0.543$& $-0.919$& $-0.328$& $-0.177$ & $-6.157$ & $42.055$ \\  
               & $-0.091$ & $0.534$ & $-0.751$ & $-2.562$ & $-1.714$ & $39.264$ & $9.814$ \\  \hline \hline
  $A_z$  & $-0.013$ & $-0.09$ & $1.305$ & $-1.500$ & $-15.38$ & $12.78$ & $213.55$  \\
         & $-0.010$ & $-0.089$ & $1.254$ & $-0.591$ & $-26.750$ & $0.989$ & $442.58$ \\ \hline \hline
  $A_u$  & $0.133$ & $0.639$ & $-1.143$ & $-0.007$ & $1.615$ & $-6.709$ & $17.189$ \\
         & $0.131$ & $0.638$ & $-1.186$ & $-3.721$ & $17.201$ & $66.602$ & $-284.922$  \\  \hline \hline
  \end{tabular}
  \end{center}
  \caption{The values for the interpolation parameters defined in eqn.\ref{interpolation-asymmetry} at LO (first 
  rows) and NLO (second rows) using full information at the partonic level.}
  \label{table-interp1}
 \end{table}

 \begin{table}[!h]
  \begin{center}
  \begin{tabular}{ c | c | c | c | c | c | c | c }
  \hline \hline
  $A_\mathcal{O}$     &  $\zeta_0$ & $\zeta_1$ & $\zeta_2$ & $\zeta_3$ & $\zeta_4$ & $\zeta_5$ & $\zeta_6$ \\ \hline \hline
  $A_{\theta_\ell^0}$ & $0.347$ & $0.628$ & $-2.062$ & $0.0126$ & $-0.026$ & $-20.652$ & $69.380$ \\  
                      & $0.297$ & $0.651$ & $-1.633$ & $-1.315$ & $-5.174$ & $4.389$ & $63.932$ \\  \hline \hline     
  $A_{x_\ell^0}$ & $-0.178$ & $0.536$ & $-0.608$ & $1.205$ & $23.422$ & $-17.264$ & $-381.304$ \\ 
                 & $-0.178$ & $0.584$ & $-0.451$ & $-2.959$ & $27.175$ & $54.492$ & $-516.980$ \\ \hline \hline
  $A_{x_\ell}$ & $-0.089$ & $0.628$& $-2.176$ &  $-1.906$ & $21.880$ & $21.691$ & $-252.617$ \\ 
               & $-0.119$ & $0.686$ & $-1.811$ & $-6.792$ & $17.724$ & $111.140$ & $-280.239$ \\  \hline \hline
  $A_z$  & $-0.090$ & $-0.150$ & $1.803$ & $-1.320$ & $-25.006$ & $-13.460$ & $334.648$  \\
         & $-0.075$ & $0.0104$ & $1.229$ & $4.397$ & $4.690$ & $-108.563$ & $-75.249$ \\ \hline \hline
  $A_u$  & $0.271$ & $0.481$ & $-2.098$ & $0.169$ & $15.691$ & $3.567$ & $-219.977$  \\  
         & $0.242$ & $0.574$ & $-1.825$ & $-5.152$ & $11.531$ & $102.039$ & $-234.340$ \\ \hline \hline
  \end{tabular}
  \end{center}
  \caption{The values for the interpolation parameters defined in eqn.\ref{interpolation-asymmetry}
  at LO (first 
  rows) and NLO (second rows) using events with kinematical cuts at the particle level.}
  \label{table-interp2}
 \end{table}

From energy and angular based observables, appropriate asymmetries are constructed.
We have generated MC samples for each value of the anomalous coupling 
$g_R$ corresponding to an integrated
luminosity of $100 \text{ fb}^{-1}$. The asymmetries were computed for each 
value of the anomalous coupling 
$$
g_R \in \{-0.2, -0.15, -0.1, 0.0, 0.1, 0.15, 0.2\}.
$$
Where we have investigated the asymmetries both at LO and NLO both at the 
parton level (without cuts) and at the particle level (with the cuts outlined in 
section \ref{sec:setup}). To model the behaviour 
of the asymmetries as function of the anomalous coupling, an 
interpolation to the computed asymmetries 
was performed. We have adopted a $6$th order polynomial defined as\footnote{Other functional
forms of the interpolation are possible as well and yield similar results.}
\begin{eqnarray}
 A_\mathcal{O} = \sum_{i=0}^6 \zeta_i^\mathcal{O} g_R^i,
 \label{interpolation-asymmetry}
\end{eqnarray}
where $\zeta_i^\mathcal{O}, i=0,\cdots,7$ is a set of coefficients
determined from the fit and corresponding to the observable $\mathcal{O}$ such that
$\zeta_0^\mathcal{O} = A_\mathcal{O}(\textrm{SM})$. In Tables \ref{table-interp1}-\ref{table-interp2}, 
we show the interpolations' results for the different 
asymmetries. 
\bibliographystyle{JHEP}
\bibliography{main.bib}

%%%%%%%%%%%%%%%%%%%%%%%%%%%%

\end{document}